\documentclass[12pt,a4paper]{article}

\usepackage{graphicx}

\input epsf

\newcommand{\frat}[2]{\frac{\textstyle #1}{\textstyle #2}}

\newcommand{\vf}[1]{\mbox{\boldmath $#1$}}

\begin{document}
\begin{center} {\Large \bf
Instanton in Euclidean non-abelian field of point-like source}\\
\vspace{0.5cm} S.V. Molodtsov, G.M. Zinovjev$^\dagger$\\
\vspace{0.5cm} {\small\it State Research Center,
Institute of Theoretical and Experimental Physics,
117259, Moscow, RUSSIA}\\
$^\dagger$ {\small \it Bogolyubov Institute for Theoretical Physics,\\
National Academy of Sciences of Ukraine, UA-03143, Kiev, UKRAINE}
\end{center}
\vspace{0.5cm}
\begin{center}
\begin{tabular}{p{16cm}}
{\small{(Anti-)instanton behaviour in Euclidean non-abelian field of
the point-like source is studied by analyzing the possible
(anti-)instanton deformations as resulted
from the variations of its characteristic parameters.
The variational principle for searching such crumpled
topological configurations is formulated and the problem
is resolved then algebraically by the Ritz method
(multipole expansion of the deformation fields).
The region of (anti-)instanton parameters relevant to the instanton
liquid model is investigated in detail. The approximate efficacious method
to include such configurations contributing to the appropriate partition
function is proposed and verified. Evaluating average energy of Euclidean
non-abelian point-like source in the instanton environment (liquid) is
performed within the superposition ansatz for the corresponding solutions
of the Yang-Mills equations. For the colour singlet dipole this energy
escalates linearly with the separation increasing and its tension
coefficient develops the magnitude commensurate
with that as inferred from lattice QCD.}}
\end{tabular}
\end{center}
\vspace{0.5cm}

\section*{Introduction}
Considerable interest in classical theory of non-abelian gauge
fields has been provoked by the discovery of instantons when it
becomes clear that conceptually low energy QCD
shares many common features with condensed matter physics.
These pseudo-particle ($PP$)
solutions of Euclidean classical equations of the Yang-Mills theory,
having been properly included in the Euclidean functional integral,
turned out very suggestive in illuminating the nature of QCD vacuum
and non-perturbative dynamics \cite{4} (the CDG scenario).
In particular, it has been realized
the QCD ground state is a sophisticated strongly interacting system filled
with the condensates of quark-antiquark pairs and gluons. Moreover,
experimentally observed particles respect only a part of the full
symmetry of QCD.
Then similarly to the condensed matter physics we have chiefly two theoretical
approaches to explore these systems. One is less theoretical and based on the
numerical simulations. Another relies on the construction of effective
theories for the low energy degrees of freedom governed by studying
the patterns of
breaking of underlying symmetries. The most advanced example is given by the
chiral symmetry of QCD in the limit of massless quarks. This symmetry is
spontaneously broken by the condensation of quarks and anti-quarks in the
QCD vacuum. Such a phenomenon (together with many consequences)
is explained to a large extent if one surmises the classical solutions
of Yang-Mills equations with nontrivial topology ((anti-)instantons with
perturbative oscillations about them added) are playing dominant role in
formation of the QCD vacuum. According to this scenario in the low energy
region the coupling constant becomes frozen somewhere at the scale of average
instanton size and the liquid (or gas) of (anti-)instantons randomly oriented
in colour space and homogeneously
distributed over 4-dimensional Euclidean space is an adequate ensemble
to saturate the corresponding functional integral. The quarks are living in
this background and its influence after all defines the observable
quantities such as the gluon and chiral condensates, the dynamically generated
quark mass the pion decay constant etc.

The pressing need in the profound study
of quasi-classical approach at calculating the generating functional of QCD
comes also from the recent developments of investigating a colour
superconductivity which appears as a new phase of hadronic matter at very
high density of quark/baryonic matter and low temperatures \cite{1}.
The analysis of quark condensate behaviour in the instanton liquid model (IL)
\cite{2} shows the corresponding phase transition takes place at a
surprisingly small value of the critical density of quark/baryonic
matter (order (or even less)
of the quark density in normal nuclei). Applying the IL scheme devised in
\cite{3} which takes into account the medium collective excitations
essentially allows to increase the critical density. Moreover, this
density of phase transition may become so high that the Coulomb
fields generated by quarks turn out comparable
to the instanton ones due to very small interquark distances.
Obviously, it puts in doubt the predictions of IL model which
ignores the influence of external fields on the pseudo-particles
and makes IL hardly applicable under such conditions.

Here we focus on studying the (anti-)instanton behaviour in the field of
point-like Euclidean colour charge. Actually, this problem is one of the
clue elements in constructing a more realistic version of IL, in
particular, at high quark/baryonic densities. Besides, it can shed
light on the challenging problem of colour field
penetration into IL. We argue and deal with the field theory approach
in quasi-classical approximation aiming to estimate the leading
contributions. Apart from scrutinizing the terms of interaction
between source and pseudo-particle which have been explored
in the pioneering papers \cite{4} and where its dipole nature has been argued,
we are trying to clarify what happens to a pseudo-particle itself when
an external field is available. Apparently, the problem of
investigating the deformation fields is of great importance in this context.
In spite of the rather impressive record of studying the pseudoparticle
interactions (see, for example, \cite{5}, \cite{51}) nowadays we are still
far from declaring this problem resolved and transformed to the practical
instrument. We formulate a further new approach to study the interactions
of point-like solutions of field equations and
suggest a new efficient way of approximate calculation of the functional
integral for interacting pseudo-particles. In this way we construct the
perturbative approach intrinsically related to the nontrivial topological
solution itself analyzing the
variations of solution parameters. In the case of pseudo-particles these
configurations retain the bulk topological features and are sensitive to
the presence of perturbative factors.  At first sight, this rather routine
excercise exhibits unexpectedly a
wealth of possibilities and turns out very intersting even out of the IL
context. We find out the pseudo-particle could be considered as a 'supplier'
of various fields, in particular, a scalar field and colour vector and
tensor fields. We expect the scalar field could be a good pretender to be
an interaction carrier in the soft momentum region \cite{6} whereas the
vector field carrying the colour indecies could be
responsible for the screening in the instanton medium.
The paper is organized as follows. The motivation to consider the
problem and supposed method of its treatment are discussed
in Sec.1 The deliberation of deformation contributions to the total
action of point-like source and instanton is given together with an
approximate approach to calculate the pseudoparticle deformations
in Sec.2 and the practical realization of this suggestion
is demonstrated in Sec.3. In the fourth Section we present the
analysis of numerical calculations and explore the generalized
formulation of searching the deformations with
imposed constraints. And eventually Section 5 is devoted
to an analysis of the results
obtained and their applications for calculating within the IL approach.

\section{Optimizing the instanton configurations}
Let us suppose the presence of immovable Euclidean colour point-like
source of external
field at the point ${\vf z}_e$. Its intensity is supposed to be ${\vf e}$
and an orientation along the third axis of colour (isotopic) space because
for the sake of simplicity we limit ourselves here to dealing with $SU(2)$
group only. Then the current density takes the form
$j^a_{\mu}=({\vf 0},~e~\delta^{a3} \delta({\vf x} -{\vf z}_e))$
where ${\vf x}$, ${\vf z}_e$ are the 3-vectors
$\mu=1,2,3,4$ (as other Greek indecies) and $a=1,2,3$
(as other Latin indecies). Fixing the location of point-like sources is gauge
invariant unlike specifing their orientation in isotopic space. In order to
avoid such
a stain \cite{6a} the static solutions of the Yang-Mills equations should be
characterized by their energy and 'total' isospin because these quantities are gauge
invariant in addition to being conserved. As known, the field
originated by the particle source in 4d-space develops the cone-like shape
with the edge being situated just at the particle creation.
Then getting away from that point the field
becomes fully developed in the area neighbouring it. In distant area
where the field penetrates into the vacuum it should be described by
the retarded solutions obeying,
in particular, the Lorentz gauge. Here we are interested in studying
the interaction in the neighbouring field area and the cylinder-symmetric
Coulomb field might be its relevant
image in 4d Euclidean space. Certainly, we skip any impact of the
instanton field on the source of colour field although the
self-consistent solution of the problem with non-abelian
fields teaches about the possible change of the source
orientation in colour space (see an example of two point-like
colour charges \cite{7}), see also \cite{71}. However, bearing in mind that
the instanton field is rather short range (the order of
the $PP$ size) we believe the source field is not reconstructed
essentially on that scale (albeit at the end of this paper we
will be well prepared to formulate the corresponding generalized
equations for the field of point-like source).

The action for gauge field $A^a_{\mu}$ when an external source is
available reads
\begin{equation}
\label{1}
S=\int dx \left( \frat14~ G_{\mu\nu}^a~G_{\mu\nu}^a+
j^a_{\mu}{\cal A}^a_{\mu}\right)~,
\end{equation}
with the field strength defined in standard way
\begin{equation}
\label{2}
G_{\mu\nu}^a=\partial_\mu {\cal A}^a_{\nu}-\partial_\nu {\cal A}^a_{\mu}+
g~\varepsilon^{abc}{\cal A}^b_{\mu}{\cal A}^c_{\nu}~,
\end{equation}
where $\varepsilon^{abc}$ is an entirely antisymmetric tensor and $g$ a
coupling constant for non-abelian field. Actually, the solution of the
Yang-Mills equations corresponding to a single (anti-)instanton, for
example, in the singular gauge is well defined and characterized by
the coordinate of the $PP$ centre $z$, the $PP$ orientation in the colour
space $\omega^{ab}$ and its size $\rho$, i.e.
\begin{equation}
\label{3}
A^a_{\mu}(x)=\frat2g~\omega^{ab}\bar\eta_{b\mu\nu}~
\frat{\rho^2}{y^2+\rho^2}~\frat{y_\nu}{y^2}~,
\end{equation}
where $y=x-z$ and $\bar\eta_{b\mu\nu}$ is the 't Hooft symbol
(for anti-instanton $\bar\eta \to \eta$). When the distance between
the point-like source and the instanton centre is large comparing
to its size $\Delta \gg \rho$ ($\Delta=|{\vf \Delta}|$,
${\vf \Delta}={\vf z}-{\vf z}_e$) the standard superposition
\begin{equation}
\label{an}
{\cal A}^a_{\mu}=B^a_{\mu}(x)+A^a_{\mu}(x)~,
\end{equation}
of point-like source field which for the sake of simplicity is
considered in this paper in the simplest gauge only
\begin{equation}
\label{4}
B^a_{\mu}(x)=({\vf 0},\delta^{a3}~\varphi),~\varphi=
\frat{e}{4\pi}~\frat{1}{|{\vf x}-{\vf z}_e|}~,
\end{equation}
and the (anti-)instanton field Eq.(\ref{3}) is a pretty reasonable
approximation of the true solution. The Lorentz gauge condition
$\partial_\mu B^a_{\mu}=0$ is valid for the retarded
solution (as for the instanton field). In the region of a well
developed field  this condition is obeyed by the Coulomb solution
Eq.(\ref{4}) as well. We treat the potentials in their Euclidean
forms but the following changes of the field and the charge of point-like
source $B_0\to iB_4$, $e\to -i e$ should be done at transition from the
Minkowski space. Actually, last variable change is resulted from the
corresponding transformations of spinor fields $\psi\to\hat\psi$,
$\bar\psi\to-i\hat\psi^\dagger$, $\gamma_0\to\gamma_4$ where the
hatted spinors are Euclidean.  Clearly, these Euclidean sources generate the
fields of the same nature as ones we face taking into account gluon
field quantum fluctuations $a_{qu}$ around the classical instanton
solutions $A=A_{inst}+a_{qu}$.

Searching the optimal $PP$ configurations we fix its size and
orientation in the isotopic space at a large distance from the source.
But approaching the source (in the local vicinity of it) both are able
to vary as the functions of the $PP$ centre coordinate, i.e. $\rho\to R(x,z)$,
$\omega^{ab}\to\Omega^{ab}(x,z)$. Now with such substitutions in the
superposition ansatz Eq.(\ref{an}) we may formulate the variation
problem of searching the deformation fields $R(x,z)$, $\Omega(x,z)$
which optimize the action Eq.(\ref{1}) with the boundary conditions
for these fields falling down at $\Delta\to\infty$. Finally, as an
output we come to the sophisticated system of differential equations which
can be analyzed numerically. However, such a scheme even treated
perturbatively faces the principle difficulty because the coefficients
of higher order derivatives of the deformation fields $R(x,z)$ and
$\Omega(x,z)$ are strongly suppressed beyond $PP$ owing to the almost
singular behaviour of the (anti-)instanton solution. This fact
essentially complicates the process of searching the appropriate
solutions and, moreover, makes it practically impossible
sometimes. At the same time this peculiarity of the $PP$ solution
behaviour prompts very natural approach to circumvent the difficulty
mentioned. It implies applying the variational
Ritz procedure (instead of the grid method) which in this
particular case can be realized as the multipole expansion of
the (anti-)instanton size, for example,
\begin{eqnarray}
\label{5}
R_{in}(x,z)&=&\rho+c_\mu~y_\mu+c_{\mu\nu}~y_\mu~ y_\nu+
\dots~,~~~~~|y|\leq L \nonumber\\
[-.2cm]
\\[-.25cm]
R_{out}(x,z)&=&\rho+d_\mu~\frat{y_\mu}{y^2}+d_{\mu\nu}~
\frat{y_\mu}{y^2}~\frat{y_\nu}{y^2}+\dots~,~~~|y|>L~. \nonumber
\end{eqnarray}
Similar expressions could be written down for the (anti-)instanton
orientation in the colour space $\Omega(x,z)$ where $L$ is a certain
parameter fixing the radius of sphere where the multipole expansion
growing with the distance increasing should be changed for the decreasing
one being a result of imposed deformation regularity constraint. Hoping
to rely further on the perturbative analysis we have to calculate the
coefficients $c_\mu,c_{\mu\nu},\dots$ and $d_\mu,d_{\mu\nu},\dots$ and
the parameter $L$ by minimizing the corresponding quadratic form
resulting from Eq.(\ref{1}), i.e. $\delta S=0$, (being calculated up
to the second order terms) together with the boundary conditions at
infinity and the conditions of sewing $'in'$ and $'out'$ parameters
on the sphere $|y|=L$. The final results allow us to conclude that
the instanton field turns out to be so strong regularizator that we
are allowed to deal with the fields $R_{in}$ and $\Omega_{in}$ only
at exploring the terms of the second order in small deviations from
$\rho$ and $\omega$. One may neglect all the specifications coming
from $R_{out}$ and $\Omega_{out}$ which means formally to operate
in the limit $L\gg\rho$.
If one is interested in stationary cylindrically-symmetric picture in
the 4-dimensional space the dependence on temporal coordinate $x_4$
should be excluded, i.e. the spatial indecies $1,2,3$ have to be
retained in the coefficients of multipole expansion Eq.(\ref{5}).
Then it is clear the final result becomes dependent on the distance
between three dimensinal components of the instanton centre
and the source position which is denoted as $\Delta$
hoping it does not produce a mess.
We would like to notice here that our condition to  defined the deformations
$\delta S=0$ allows us to fix the major contributions to the exponent of the
generating functional. In a sence we might consider this equation as the valley
equation in the quasiclassical approximation. Then the higer order terms should
reproduce loop corrections (would be quantum interactions).

Going to implement the program mentioned above let us
consider the example of the dipole expansion terms (\ref{5}) denoting them as
\begin{equation}
\label{6}
\delta\rho=\delta_\mu\rho~y_\mu~,~~~\delta\omega=\delta_\mu \omega~y_\mu~,
~~~R=\rho+\delta\rho~,~~~\Omega=\omega+\delta\omega~.
\end{equation}
Then for the crumpled instanton we have instead of (\ref{3}) the following
expansion in $\delta\rho$ up to the second order terms
$$A^a_{\mu}\simeq A^a_{\mu}(\Omega,\rho)+
\frat{\partial A^a_{\mu}(\Omega,\rho)}
{\partial R}~\Delta R+ \frat{\partial^2 A^a_{\mu}(\Omega,\rho)}
{\partial R~\partial R}~\frat{(\Delta R)^2}{2}~, $$
here $\Delta R=R-\rho$. Using the superposition $B^a_{\mu}+A^a_{\mu}$ in
Eq.(\ref{2}) leads to the definition of $G_{\mu\nu}^a$ as
\begin{eqnarray}
\label{7}
&&G_{\mu\nu}^a=G_{\mu\nu}^a(B)+G_{\mu\nu}^a(A)+G_{\mu\nu}^a(A,B)~,\nonumber\\
[-.2cm]
\\[-.25cm]
&&G_{\mu\nu}^a(A,B)=g~\varepsilon^{abc} (B^b_{\mu}A^{c}_\nu+
A^b_{\mu}B^{c}_\nu )~,\nonumber
\end{eqnarray}
with $G_{\mu\nu}^a(A)$ and $G_{\mu\nu}^a(B)$ in the standard form of
Eq.(\ref{2}) and the field strength squared looks then like the following sum
\begin{eqnarray}
\label{8}
G_{\mu\nu}^a~ G_{\mu\nu}^a &=&G_{\mu\nu}^a(B)~G_{\mu\nu}^a(B) +
G_{\mu\nu}^a(A)~G_{\mu\nu}^a(A) +G_{\mu\nu}^a(A,B)~G_{\mu\nu}^a(A,B)
+\nonumber\\
[-.2cm]
\\[-.25cm]
&+&2~ G_{\mu\nu}^a(B)~G_{\mu\nu}^a(A) + 2~ G_{\mu\nu}^a(B)~G_{\mu\nu}^a(A,B)+
2~G_{\mu\nu}^a(A)~G_{\mu\nu}^a(A,B)~.\nonumber
\end{eqnarray}
One should not envisage this complicated expression tractable in the
approximation of weak external field because of the strongly singular
nature of instanton solution and point-like source field. Besides,
the large number of interacting terms essentially contributing does
not allow one to be based on their physical meaning only (as in
\cite{pol}) in further analysis.  Now let us note the
non-zero strength component of point-like source field reads
$$G^a_{4i}(B)=\frat{e}{4\pi}~\delta^{a3}~\frat{y_i+
\Delta_i}{|{\vf y}+ {\vf \Delta}|^3}~.$$
whereas the field strength generated by the instanton is presented as
\begin{equation}
\label{strng} G_{\mu\nu}^a(A)=\hat G_{\mu\nu}^a(A)+\Delta G_{\mu\nu}^a(A)~,
\end{equation}
and the first term here looks like instanton field strength but with the
varying parameters
$$\hat G_{\mu\nu}^a(A)=-\frat{4}{g}~\Omega^{ak} M_{\mu\alpha} M_{\nu\beta}~
\bar\eta_{k\alpha\beta}~\frat{R^2}{(y^2+R^2)^2}~,$$
where $M_{\mu\alpha}=\delta_{\mu\alpha}-2~\frat{y_\mu y_\alpha}{y^2}$,
and the correction term is given by
\begin{eqnarray}
&&\Delta G_{\mu\nu}^a(A)=\frat{\partial A^a_{\nu}(\Omega,R)}
{\partial R}~\frat{\partial R}
{\partial y_\mu} -\frat{\partial A^a_{\mu}(\Omega,R)}
{\partial R}~\frat{\partial R}
{\partial y_\nu} +\frat{\partial A^a_{\nu}(\Omega,R)}{\partial \Omega^{bc}}~
\frat{\partial \Omega^{bc}}{\partial y_\mu} -\frat{\partial
A^a_{\mu}(\Omega,R)}
{\partial \Omega^{bc}}~\frat{\partial \Omega^{bc}}{\partial y_\nu}+\nonumber\\
&&+\frat{4}{g}[\varepsilon^{abc}(\omega^{bm} \delta\omega^{cn}-
\omega^{bn}\delta
\omega^{cm}+\delta \omega^{bm}\omega^{cn})-\delta
\omega^{ak}\varepsilon^{kmn}]~
\bar\eta_{m\mu\alpha}\bar\eta_{n\nu\beta}~\frat{R^4}{(y^2+R^2)^2}
~\frat{y_\alpha y_\beta}{y^4}~.\nonumber
\end{eqnarray}
Finally, the interfering term takes the following form
$$G_{4i}^a(A,B)=2~ \frat{e}{4\pi}~\varepsilon^{a3c}~
\Omega^{ck}~\bar\eta_{ki\alpha}~
\frat{y_\alpha}{y^2}~\frat{R^2}{(y^2+R^2)}~ \frat{1}{|{\vf y}+
{\vf \Delta}|}~.$$
The other terms do not contribute to $G^a_{\mu\nu}(A,B)$ because
of the particular choice of gauge. Later we will develop the
perturbative approach for this part of strength expanding it
in $\delta\rho$ ш $\delta\omega$ up to the second order terms.

\section{Classifying the deformation contributions}
The first term of Eq.(\ref{8}) is certainly out of our interest here and
calculating the action with the second term included gives the correction
to single instanton action. The quadratic form of kinetic energy type
hinges upon the deformation fields
\begin{eqnarray}
\label{9}
&&S_{kin}=\int dx~ \frat14~G_{\mu\nu}^a(A)G_{\mu\nu}^a(A)-\beta=
\frat{\kappa}{2}~
(\delta_\mu\rho)^2+ \frat{\sigma_{\mu\nu}^{kl}}{2}~
\delta_\mu\omega^{ak}\delta_\nu\omega^{al}+\frat{\sigma_3}{2}~
\varepsilon^{kmn}\varepsilon^{abc} \omega^{bm}\delta_\mu \omega^{ak}
\delta_\mu \omega^{cn} +\nonumber\\
[-.2cm]
\\[-.25cm]
&&+\frat{\sigma_4}{2}~\bar\eta_{m\mu\nu}\varepsilon^{abc}
(\omega^{bm}\delta_\nu \omega^{cn}-\omega^{bn}\delta_\nu \omega^{cm})
\delta_\mu \omega^{an}+\frat{\sigma_5}{2}
~\omega^{am} \omega^{cn}\delta_\mu \omega^{am}\delta_\mu \omega^{cn}
+\widetilde v^{al}_{\mu\nu}~  \delta_\mu\omega^{al}\delta_\nu\rho ,\nonumber
\end{eqnarray}
with the tensor coefficients
\begin{eqnarray}
\label{kin}
&&\kappa=\frat{9}{10}~\beta~,~~ \sigma_{\mu\nu}^{kl}=\sigma_1~\delta_{\mu\nu}
\delta_{kl}+\sigma_2~\varepsilon_{kln}~\bar\eta_{n\mu\nu}~,~~
\widetilde v^{al}_{\mu\nu}=\omega^{ak}v^{kl}_{\mu\nu}~,
~~v^{kl}_{\mu\nu}=v_1~\delta_{\mu\nu}
\delta_{kl}+v_2~\varepsilon_{kln}~\bar\eta_{n\mu\nu}~, \nonumber\\
&&\sigma_1=\frat{23}{24}~\beta~\rho^2~,
~~\sigma_2=-\frat{3}{8}~\beta~\rho^2~, ~~\sigma_3=-2v_1~\rho~,\\
&&\sigma_4=2v_2~\rho~,~~\sigma_5=\frat{4}{3}~v_2~\rho~,
~~v_1=\frat{7}{24}~\beta~\rho~,~~v_2=\frat{1}{8}~\beta~\rho~,\nonumber
\end{eqnarray}
where $\beta=\frat{8\pi^2}{g^2}$ is the single (anti-)instanton action.
These contributions are regulated by instanton itself and independent
of the field $B$ as well as the distance between the pseudo-particle
center and the source.
In a sense they characterize the $PP$ compliantness as for its shape
changes and an orientation in colour space. Here is an appropriate
place to notice that dealing with
the crumpled configurations characterized by the parameters (\ref{6})
within the perturbative approximation and in the limit $L\to\infty$
(see (\ref{5})) presupposes a lot of singular terms available.
Fortunately, the natural regularization pops
in when those bad terms are multiplied by the instanton tensor
$G^a_{\mu\nu}(\rho,\omega)$ (see below (\ref{10d})).
It is interesting to notice the operator similar to Eq.(\ref{9})
emerges when one explores the fluctuation fields in the proximity
of the particular classical solution $A=A_{cl}+a_{qu}$ and the
action acquires an additional quadratic term
\begin{equation}
\label{Laq}
S=S(A_{cl})+\int dx~ a_{qu}~ {\cal L}(A_{cl})~ a_{qu}=S(A_{cl})+
\sum_{k} \lambda_k ~\xi_k^{2}~,
\end{equation}
where $\lambda_k$ are the eigenvalues of the operator ${\cal L}$
and $\xi_k$ presents the coefficients of the expansion of field
$a_{qu}=\sum_{k} \xi_k \psi_k$ into the entire set of
orthogonal eigenfunctions $\psi_k$ of this operator. In view of that
generally Eq.(\ref{9}) delineates the parametric dependence of
eigenvalues and eigenfunctions of the operator ${\cal L}$ in the
accepted approximation. In our particular case it means
dependence on the expansion coefficients in Eq.(\ref{5})
$\left(\frat{\partial \lambda_k}{\partial c_\mu},
\frat{\partial \psi_k}{\partial c_\mu},\dots\right)$.
Apparently, we gain the possibility to take into account the
contribution of regular component originated by external field in
the functional integral instead of dealing with poorly defined
Green function of instanton.

\begin{figure*}
[!tbh]
\begin{center}
\includegraphics[width=0.75\textwidth]{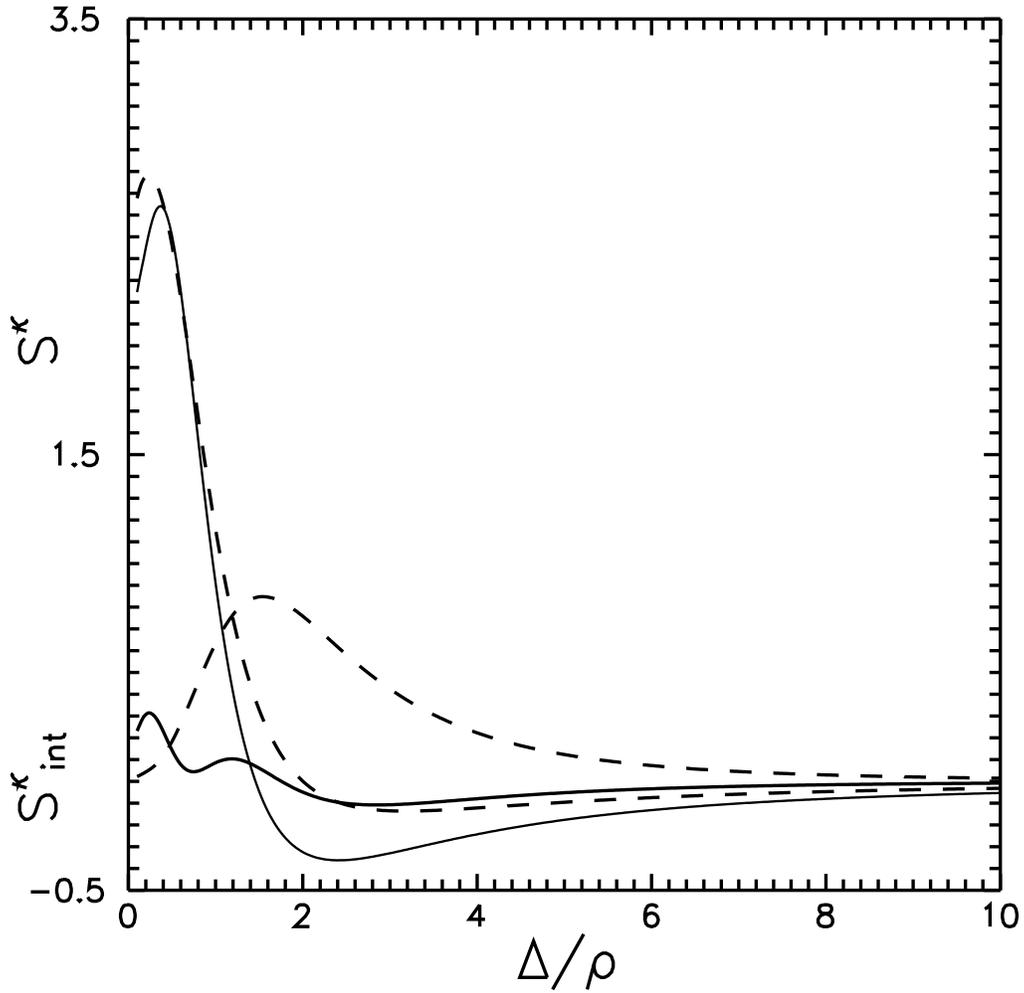}
\end{center}
\vspace{-7mm}
\caption{The comparison of contributions $S_{kin}$ and $S^{\kappa}_{int}$
as the functions of the path along $x$ axis. The instanton action is
$\beta=18$ and the source intensity is taken as $e=g$. The rotation matrix
$\omega^{ak}$ is defined by the parameters ${\vf n}=(0,1,0)$,
$\phi=\pi/2$ (see Eq.(\ref{13})). Two upper curves (in the vicinity
of coordinate origin) demonstrate the $S_{kin}$ behaviour
whereas two lower ones demonstrate the $S_{int}^\kappa$ behaviour. The solid
curves correspond to the instanton and the dashed ones correspond to the
anti-instanton. The source $e$ is oriented along the third axis in the
isotopic space.}
\label{skapin}
\end{figure*}

Calculating Eq.(\ref{kin}) and some subsequent ones we used the
following identities valid for $SU(2)$-group
\begin{equation}
\label{ort}
\omega^{ab}\omega^{ac}=\delta^{bc}~,~~
\varepsilon^{abc}\omega^{a\alpha}\omega^{b\beta}\omega^{c\gamma}=
\varepsilon^{\alpha\beta\gamma}~.\nonumber
\end{equation}

It is practical for analysing the additional term to the single
instanton action to present it as a sum of two components
\begin{equation}
\label{10}
S-\beta=S_{kin}+S_{int}~.
\end{equation}
where $S_{kin}$ is the term of kinetic energy type and $S_{int}$ is
the interacting term (of course, the singular contribution of the
point-like source to the action is ignored). The interacting term
$S_{int}$ may be decomposed into the following sum
\begin{equation}
\label{10a}
S_{int}=S_{int}^\kappa+S^d_{int}+S^\delta_{int}~.
\end{equation}
and the physical meaning of each term here becomes clear from further
calculations. We quote here the final result only ommitting the routine
calculations which are not very inventive and require the good patience
mainly.

\subsection{Quadratic terms depending on the distance from source}
As the first term of Eq.(\ref{10a}) we mean the quadratic terms in
$\delta\rho$, $\delta\omega$ wholesale which describe the dependence
of kinetic coefficients on $\Delta$:
\begin{eqnarray}
\label{10b}
&&S^\kappa_{int}=\frat{1}{g}\frat{e}{4\pi}~\omega^{3k}\bar\eta_{k4l}~
\Sigma_{lij}~\delta_i\rho~\delta_j\rho+\left(\frat{e}{4\pi}\right)^2
(\delta_{kl}-\omega^{3k}\omega^{3l})(\delta_{kl}~Q_{ij}-\bar\eta_{k4m}
\bar\eta_{l4n}~R_{mnij})~\delta_i\rho~\delta_j\rho-\nonumber\\
&&-\frat{1}{g}\frat{e}{4\pi}~\bar\eta_{k4m}(\bar\eta_{lin}~S_{jmn}+
\bar\eta_{lmn}~T_{nij})\varepsilon^{a3b}~\delta_i\omega^{ak}~
\delta_j\omega^{bl}-\frat{1}{g}\frat{e}{4\pi}~
\varepsilon^{a3b}\bar\eta_{k4m}\bar\eta_{lnc}~Z_{mcijn}~\delta_i\omega^{ak}
\delta_j\omega^{bl}+\nonumber\\
&&+\frat{1}{g}\frat{e}{4\pi}~\varepsilon^{a3c}
\bar\eta_{kl\alpha}\bar\eta_{m4\beta} \bar\eta_{nl\gamma}~
\Phi_{\alpha\beta\gamma ji}
~[\varepsilon^{abd}(\omega^{bm}\delta_i\omega^{dn}
-\omega^{bn}\delta_i\omega^{dm})- \varepsilon^{dmn}
\delta_i\omega^{ad}]\delta_j\omega^{ck}+\nonumber\\
[-.2cm]
\\[-.25cm]
&&+\left(\frat{e}{4\pi}\right)^2(\delta_{kl}~U_{ij}-
\bar\eta_{k4m}\bar\eta_{l4n}~V_{ijmn})(\delta_i\omega^{ak}~
\delta_j\omega^{al}-\delta_i\omega^{3k}~ \delta_j\omega^{3l})+
\frat{1}{g}\frat{e}{4\pi}~\bar\eta_{k4l}~
\Psi_{ilj}~\delta_i\omega^{3k}~\delta_j\rho+\nonumber\\
&&+\frat{1}{g}\frat{e}{4\pi}[ (\bar\eta_{l4m}\bar\eta_{kmn}-
\bar\eta_{k4m}\bar\eta_{lmn})W_{ijn}+(\bar\eta_{l4m}\bar\eta_{kjn}-
\bar\eta_{k4m}\bar\eta_{ljn})X_{mni} +2\bar\eta_{l4m}\bar\eta_{kin}
X_{mnj}]\omega^{ak}\varepsilon^{a3b}\delta_i\omega^{bl}
\delta_j\rho+\nonumber\\
&&+\frat{1}{g}\frat{e}{4\pi}~\varepsilon^{a3c}\omega^{ck}
\bar\eta_{kl\alpha}\bar\eta_{m4\beta} \bar\eta_{nl\gamma}~\Theta_{\alpha i
\beta\gamma j} ~[\varepsilon^{abd}(\omega^{bm}\delta_i\omega^{dn}-
\omega^{bn}\delta_i\omega^{dm})-
\varepsilon^{dmn}\delta_i\omega^{ad}]~\delta_j \rho+\nonumber\\
&&+\frat{1}{g}\frat{e}{4\pi}~\bar\eta_{c4k}\bar\eta_{dln}
~\Xi_{knijl}~[\varepsilon^{3ab} (\omega^{ac}\delta_i\omega^{bd}-
\omega^{ad}\delta_i\omega^{bc})-
\varepsilon^{acd}\delta_i\omega^{3a}]~\delta_j \rho~.\nonumber
\end{eqnarray}
Here we are studying an impact of fully developed Coulomb field
on an instanton and that is why the above formula includes the spatial
indecies only of the deformation fields
$\delta_\mu\rho$, $\delta_\mu\omega$.
The overt but cumbersome representations of the
tensor components are given in Appendix. In this paragraph we are
interested in their asymptotic values at $\Delta\to\infty$ only and
it is convenient for further analysis to use the dimensionless
coordinates according to the following change
$\Delta_i/\rho \to \Delta_i$. The terms of $O((\delta\rho)^2)$ are
presented by three tensors. One of those is the third rank tensor
\begin{eqnarray}
\label{10c}
\Sigma_{ijk}=\delta_{ij}\hat\Delta_k~\Sigma_1+
\delta_{ik}\hat\Delta_j~\Sigma_2+\delta_{jk}\hat\Delta_i~\Sigma_3+
\hat\Delta_i\hat\Delta_j\hat\Delta_k~\Sigma_4~,\nonumber\\
\Sigma_1\simeq -2\pi^2~,~~\Sigma_2\simeq 2\pi^2~,~~
\Sigma_3\simeq \frat{49\pi^2}{3}\frat{\ln \Delta}{\Delta^2}~,~~
\Sigma_4\simeq 0~,\nonumber
\end{eqnarray}
where $\hat\Delta_i=\frat{\Delta_i}{\Delta}$ is the unit vector;
another one is the entirely symmetric tensor of the second rank
\begin{equation}
\label{t2}
Q_{ij}=\delta_{ij}~Q_1+\hat\Delta_i\hat\Delta_j~Q_2~,
\end{equation}
with the components
$$Q_1\simeq 6\pi^2\frat{\ln\Delta}{\Delta^2}~,~~Q_2\simeq 0~;$$
and eventually the entirely symmetric tensor of the fourth rank
\begin{eqnarray}
\label{t4} R_{ijkl}&=&(\delta_{ij}\delta_{kl}+\delta_{ik}\delta_{jl}+
\delta_{il}\delta_{jk})~R_1+ (\delta_{ij}\hat\Delta_k\hat\Delta_l+
\delta_{ik}\hat\Delta_j\hat\Delta_l+
\delta_{il}\hat\Delta_j\hat\Delta_k+ \nonumber\\
 [-.2cm]
\\[-.25cm]
&+&\delta_{kl}\hat\Delta_i\hat\Delta_j+\delta_{jl}\hat\Delta_i\hat\Delta_k+
\delta_{jk}\hat\Delta_i\hat\Delta_l)~R_2+
\hat\Delta_i\hat\Delta_j\hat\Delta_k\hat\Delta_l~R_3,\nonumber
\end{eqnarray}
where
$$R_1\simeq\pi^2\frat{\ln\Delta}{\Delta^2}~,~~R_2\simeq 0~,~~R_3\simeq 0~.$$
If the particular component of any tensor is small comparing to the leading
asymptotic term and is out of discussion we imply zero value for it and
don't show its asymptotic form. All the tensor-functions, excluding the
components $\Sigma_1$, $\Sigma_2$ of $\Sigma_{ijk}$, are decreasing with
the distance increasing. An appearence of these nonzero asymptotic values at
$\Delta \to \infty$ results from the contribution of interference term
originated by the product of source field strength $G^a_{4i}(B)$ and
the $G^a_{4i}(A)$ component of second order in $(\delta\rho)^2$. Such
behaviour is generated by the substitution where the parameter fixing
above mentioned radius of multipole expansion $L\to\infty$.

The coefficients of the $O((\delta\omega)^2)$ terms are constructed by two
entirely symmetric tensors $S$ and $T$ of the third order. They are
\begin{equation}
\label{t3}
S_{ijk}=(\delta_{ij}\hat\Delta_k+\delta_{ik}\hat\Delta_j+
\delta_{jk}~\hat\Delta_i)~S_1+ \hat\Delta_i\hat\Delta_j\hat\Delta_k~S_2~,
\end{equation}
with the components
$$S_1\simeq-\frat{\pi^2}{3}\frat{\ln\Delta}{\Delta^2}~,~~ S_2\simeq 0~,$$
and the tensor {T} is analogous to Eq.(\ref{t3}) with similarly looking
components
$$T_1\simeq-\frat{\pi^2}{2}\frat{\ln\Delta}{\Delta^2}~,~~ T_2\simeq 0~.$$
Besides, Eq.(\ref{10b})contains the fifth rank tensor
$Z$ in the following form
\begin{eqnarray}
\label{t5}
Z_{ijklm}&=&[\hat\Delta_i(\delta_{jk}\delta_{lm}+
\delta_{jl}\delta_{km}+\delta_{jm}\delta_{kl})+
\hat\Delta_j(\delta_{ik}\delta_{lm}+
\delta_{il}\delta_{jm}+\delta_{im}\delta_{kl})+\nonumber\\
&+&\hat\Delta_k(\delta_{ij}\delta_{lm}+
\delta_{il}\delta_{jm}+\delta_{im}\delta_{jl})+
\hat\Delta_l(\delta_{ij}\delta_{km}+\delta_{ik}\delta_{jm}
+\delta_{im}\delta_{jk})]~Z_1+\nonumber\\
&+&\hat\Delta_m(\delta_{ij}\delta_{kl}+
\delta_{ik}\delta_{jl}+\delta_{il}\delta_{jk})~Z_2+
\hat\Delta_m[\delta_{ij}\hat\Delta_k\hat\Delta_l+
\delta_{ik}\hat\Delta_j\hat\Delta_l+ \delta_{il}\hat\Delta_j\hat\Delta_k+\\
&+&\delta_{jk}\hat\Delta_i\hat\Delta_l+
\delta_{jl}\hat\Delta_i\hat\Delta_k+
\delta_{kl}\hat\Delta_i\hat\Delta_j]~Z_3
+[\delta_{im}\hat\Delta_j\hat\Delta_k\hat\Delta_l+\nonumber\\
&+&\delta_{jm}\hat\Delta_i\hat\Delta_k\hat\Delta_l
+\delta_{km}\hat\Delta_i\hat\Delta_j\hat\Delta_l
+\delta_{lm}\hat\Delta_i\hat\Delta_j\hat\Delta_k]~Z_4
+\hat\Delta_i\hat\Delta_j\hat\Delta_k\hat\Delta_l\hat\Delta_m~Z_5 ~,\nonumber
\end{eqnarray}
their components have no singularities and are going to zero in the limit
$\Delta \to \infty$. There exists the contribution of entirely symmetric
tensor of the fifth rank $\Phi$ in Eq.(\ref{10b}) which has the form
analogous to $Z$ with the components $\Phi_1=\Phi_2$, $\Phi_3=\Phi_4$.
In the Appendix below two first components are denoted by $\Phi_1$,
another two as $\Phi_2$ and, finally, the component $\Phi_5$ is denoted
by $\Phi_3$. Unlikely the tensor $Z$ the component of tensor $\Phi$
having two time indecies ($\Phi_{4\beta4ji}$) provides the terms of
kinetic energy type with non-zero contribution. We demonstrate two
nontrivial components of construction of the $S_{ijk}$-type
(Eq.(\ref{t3})) which are denoted as $\Phi^{44}_{1,2}$ in Appendix.
As tensor $Z$ the tensor $\Phi$ does not develop the
singularities in the limit $\Delta \to \infty$. The other terms of
Eq.(\ref{10b}) are given by the entirely symmetric tensor of the second
rank $U$ similar to Eq.(\ref{t2}) with the components
$$U_1\simeq\pi^2~\frat{\ln\Delta}{\Delta^2}~,~~U_2\simeq 0~;$$
and the entirely symmetric tensor of the fourth rank $V$ of the
Eq.(\ref{t4}) type with the components
$$V_1\simeq-\frat{\pi^2}{4}\frat{\ln\Delta}{\Delta^2}~,
~~V_2\simeq 0~,~~V_3\simeq 0~.$$
In distinction with the dilatation term $\Sigma$ in the contribution of
kinetic energy type Eq.(\ref{10b}) the term generated by the rotations in the
isotopic space has no singular contributions (they compensate each other
at the stage of intermediate calculations) and in output the tensor
components are decreasing with the distance increasing.

For the mixed contributions of $O(\delta\rho~\delta\omega)$ we have the
tensor with the component approaching a constant value again
\begin{eqnarray}
\label{10g}
&&\Psi_{ijk}=(\delta_{ij}\hat\Delta_k+\delta_{jk}\hat\Delta_i)~\Psi_1+
\delta_{ik} \hat\Delta_j~\Psi_2+ \hat\Delta_i\hat\Delta_j\hat
\Delta_k~\Psi_3~,\nonumber\\
&&\Psi_1\simeq 0~,~~\Psi_2\simeq 0~,~~\Psi_3\simeq 4\pi^2~;\nonumber
\end{eqnarray}
(the singularity in the third component is related to the term $jA$ of
the action Eq.(\ref{1})), and two other entirely symmetric tensors $W$
and $X$ of the third rank (of the Eq.(\ref{t3}) type again) with the
following components
$$W_1\simeq-2\pi^2\frat{\ln\Delta}{\Delta^2}~,~~
W_2\simeq 0~,~~ X_1\simeq-\frat{4\pi^2}{3}\frat{\ln\Delta}
{\Delta^2}~,~~ X_2\simeq 0~. $$
The entirely symmetric tensor of the fifth rank $\Theta$ available in
Eq.(\ref{10b}) is constructed to be similar to $\Phi$ with the
nontrivial components $\Theta_{1,2,3}$,
$\Theta^{44}_{1,2}=\Theta_{4i\beta4j}$. And, finally, the fifth rank
tensor $\Xi$ takes the form of Eq.(\ref{t5}). Both tensors $\Theta$
and $\Xi$ have no the singularities in the limit $\Delta \to \infty$.

Now collecting the asymptotic values obtained all together we have for
the additional contribution to the kinetic energy at $\Delta\to\infty$
(within the logarithmic precision) the following equation
\begin{equation}
\label{10d}
S^\kappa_{int}\to\frat{1}{g}\frat{e}{4\pi}~2\pi^2\omega^{3k}
\bar\eta_{k4l}~(-\hat\Delta_j \delta_{li}+
\hat\Delta_i\delta_{lj})~\delta_i\rho~\delta_j\rho=0~,
\end{equation}
which implies that any singularities available in the components of
tensors do not produce any impact on the observables and, hence,
the simplified procedure of the Ritz method with the parameter
$L\to\infty$ is quite applicable.

The complete analysis of the $PP$
deformations including the contribution of the term $S^\kappa_{int}$
Eq.(\ref{10b}) may be cause for concern. However, the approximate
procedure of searching the optimal deformations could be
suggested just for the parameters interesting for applications in the IL model
($\beta=\frat{8\pi^2}{g^2}\simeq 12 \div 18$) \cite{8}.
Let us address Fig.1 where the contributions $S_{kin}$ and $S^\kappa_{int}$
for the single instanton are compared as functions of the path along the
$x$-axis. We suggest for the instanton action $\beta=18$ and the point-like
souce intensity is taken as $e=g$. As to the rotation matrix it is defined by
the vector ${\bf n}=(0,1,0)$ which fixes the $PP$ rotation in the isotpic
space to the angle $\varphi$ in the following form
\begin{equation}
\label{13}
\omega^{ab}=n_a n_b +P_{ab}\cos \varphi -\varepsilon_{abc}~n_c \sin \varphi~,
\end{equation}
where $P_{ab}=\delta_{ab}-n_a n_b$ is the appropriate projection operator.
In the example under discussion it is taken $\varphi=\pi/2$. If one neglects
the contribution of $S^\kappa_{int}$ (comparing it to the contribution of
$S_{kin}$) the deformations $\delta\rho$, $\delta\omega$ could be calculated
in the true form (see Eq.(\ref{12}) below). It is clear from Fig.1 the
application of this approximation is well justified at small
distances from the point-like source for both the instanton and
the anti-instanton. The contribution $S_{int}^\kappa$ for the anti-instanton
in the region of instanton size order is rather large formally but in this
case the contribution of the term of kinetic
energy type is small compared to $S^\delta_{int}$ and the solution may be
also used for the rough quantatative estimate. In what follows all the
figures are obtained for the same parameters of rotation matrix and
the same path as in Fig.1.

\begin{figure*}
[!tbh]
\begin{center}
\includegraphics[width=0.75\textwidth]{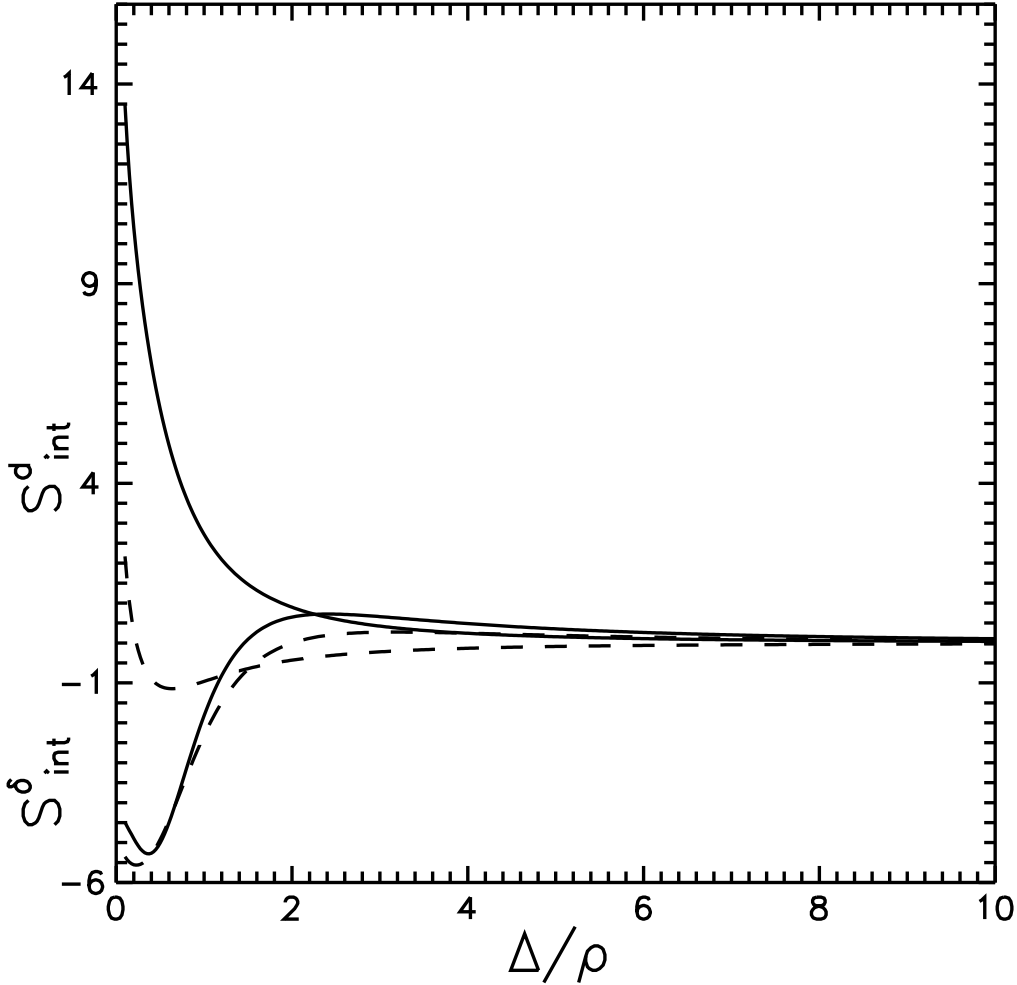}
\end{center}
\vspace{-7mm}
\caption{Simile of the contributions to $S_{int}$. The path and parameters
which define the rotation matrix are the same as in Fig. 1. Two upper
curves show $S^d_{int}$ and two lower ones demonstrate $S^{\delta}_{int}$.
The solid curves correspond to an instanton whereas both dashed
curves correspond to an anti-instanton.}
\label{sdel}
\end{figure*}

\subsection{Analysing interaction of the undeformed instanton}
Here we are going to analyze the other interaction terms in Eq.(\ref{10a}).
The second term there hinges directly on the asymptotic value of
instanton size $\rho$ and orientation $\omega^{ak}$ and describes the
interaction of undeformed instanton with the point-like source
\begin{equation}
\label{11}
S^d_{int}=\frat1g~\frat{e}{4\pi}~\bar\eta_{k4i}~\omega^{3k} I_i
+\left(\frat{e}{4\pi}\right)^2
J+ \left(\frat{e}{4\pi}\right)^2 K_{kl}~\omega^{3k}\omega^{3l}~,
\end{equation}
It is interesting to notice the first term with the coefficient
$$I_i=8~\pi^2~\Delta_i~(D^{-1}-\Delta^{-1})~$$
where $D^2=\Delta^2+1$ is entirely generated by the term $jA$ of
initial action Eq.(\ref{1}). And the contributions
of chromoelectric and chromomagnetic fields to $S^d_{int}$ in two terms
\begin{equation}
\label{11a} 2~G_{\mu\nu}^a(B)~G_{\mu\nu}^a(A) +
2~ G_{\mu\nu}^a(A)~G_{\mu\nu}^a(A,B)~,
\end{equation}
after integrating are cancelled. If it occurs the instanton field is a weak
perturbation of the Coulomb background (just the CDG approximation)
then the first term becomes dominant
and one can obtain
$$\frat14\int dx~2~G_{\mu\nu}^a(B)~G_{\mu\nu}^a(A)=
\frat{e}{4\pi}~\frat{8\pi^2}{g}~ \bar\eta_{k4i}~\omega^{3k}~
\frat{\Delta_i}{\Delta^2}~(2~D^3-2~\Delta^3 -3 D +D^{-1})~.$$
Turning to the limit $\Delta \to \infty$ we get the well known result for
the dipole interaction of an instanton with an external field
$$\frat14 \int dx ~ 2~ G_{\mu\nu}^a(B)~G_{\mu\nu}^a(A)\to \frat{2\pi^2}{g}~
\bar\eta_{k4i}~\omega^{3k}~E_i~,$$
if the transition to the Minkowski space is performed and the dimensional
variables are restored (see remarks after Eq.(\ref{4})). The vector $E_i$
above presents the source field in the following form
$$E_i=\frat{e}{4\pi}~\frat{\Delta_i}{\Delta^3}~.$$
The function
$$J=2~\int dy ~\frat{2~y^2- {\vf y}^2}{y^4~(y^2+1)^2~|{\vf y} +
{\vf \Delta}|^2}~,$$
and tensor
$$K_{kl}=2~\int dy~ \frat{y_k y_l}{y^4}~\frat{1}{(y^2+1)^2~|{\vf y}
+ {\vf \Delta}|^2}~,$$
cannot be integrated in the elementary functions although numerically
can be analysed comprehensively. Their asymptotic values at
$\Delta\to\infty$ look simply
$$J\simeq \frat{5\pi^2}{2} \frat{1}{\Delta^2}~,$$
and for the components of the second rank tensor Eq.(\ref{t2}) as
$$K_1\simeq\frat{\pi^2}{2}\frat{1}{\Delta^2}~,~~K_2\simeq 0~.$$
Now taking for the action of single instanton, for example, $\beta=18$ and
for the source intensity $e=g$ we find out the contribution of the terms
proportional $e^2$ in Eq.(\ref{11}) at small distances ($\Delta\leq 1$) is
noticeably larger than the first term contribution. The behaviour of the
curve corresponding to $S^d_{int}$ in Fig.2 allows us to make such a
conclusion. If the contribution of the first dipole-like term in Eq.(\ref{11})
is the leading one the dashed curve corresponding to the anti-instanton should
run completely in the region of negative values. Further on in this paper we
face another interesting result related to the slow decreasing of these
functions while applied to IL.

\subsection{Linear deformations}
Apparently, the interaction terms which allow us to detach the optimal
deformations of $PP$ are of major interest in this paper. Those terms are
just linear in $\delta\rho$ and $\delta\omega$ and could be given in
the following form
\begin{equation}
\label{lm}
S^\delta_{int}=\Lambda_n~\delta_n\rho+M^{ak}_n~ \delta_n \omega^{ak}~,
\end{equation}
where
\begin{equation}
\label{Ln}
\Lambda_n=\frat1g~\frat{e}{4\pi}~\bar\eta_{k4i}~
\omega^{3k}~A_{in}+\left(\frat{e}{4\pi}\right)^2~B_n+
 \left(\frat{e}{4\pi}\right)^2~\omega^{3k}\omega^{3l}~C_{knl}~.
\end{equation}
The tensor components are cited in Appendix and here we give their
asymptotic values at large distances. For example, for the second
rank tensor $A$ which is similar Eq.(\ref{t2}) these
values appear as
$$ A_1\simeq-2\pi^2\frat{1}{\Delta}~,~~A_2\simeq16\pi^2\frat{1}{\Delta}~.$$
The vector-function $B_i$ spanned on the unit vector $\hat\Delta$ has the form
$$B_i=\hat\Delta_i~B~,~~B\simeq-\frat{28\pi^2}{3}\frat{\ln\Delta}{\Delta^3}~$$
and two components of the third rank tensor $C$ similar Eq.(\ref{t3}) are
$$C_1\simeq-\frat{4\pi^2}{3}\frat{\ln\Delta}{\Delta^3}~,~~ C_2\simeq 0~.$$
Finally, the term in Eq.(\ref{lm}) related to the rotations in isotopic
space is defined by the following tensor-functions
\begin{eqnarray}
\label{Mn}
M_n^{ak}&=&\frat1g~\frat{e}{4\pi}\bar\eta_{k4i}~D_{in}~\delta^{a3}-
\frat1g~\frat{e}{4\pi}~\varepsilon^{a3c}~\omega^{cl}(\bar\eta_{k4m}
\bar\eta_{lni}~E_{mi} +\bar\eta_{k4j}\bar\eta_{ljm}~F_{mn})+\nonumber\\
&+&\frat1g~\frat{e}{4\pi}~[\varepsilon^{a3b}~\omega^{bc}(\bar\eta_{c4j}
\bar\eta_{kil}-\bar\eta_{k4j}\bar\eta_{cil})-\delta^{a3}
\varepsilon^{kbc}\bar\eta_{b4j}\bar\eta_{cil}]~H_{jlni}+\nonumber\\
[-.2cm]
\\[-.25cm]
&+&\frat1g~\frat{e}{4\pi}~[(\omega^{ab}\omega^{3c}-\delta^{3a}
\delta^{bc})\delta^{kl}-(\omega^{ab}\omega^{3l}-\delta^{3a}
\delta^{lb})\delta^{kc}-\varepsilon^{a3d}\omega^{db}\varepsilon^{kcl}]
~\bar\eta_{bi\alpha}\bar\eta_{c4\gamma}\bar\eta_{li\delta}~
Y_{\alpha n \gamma \delta}+\nonumber\\
&+&\left(\frat{e}{4\pi}\right)^2(\omega^{al}-\omega^{3l}\delta^{a3})
 (O_n\delta_{kl}-P_{lkn})~.\nonumber
\end{eqnarray}
Then the corresponding asymptotic values when $\Delta\to\infty$ for two
components of the $D$ tensor similar to Eq.(\ref{t3}) are
$$D_1\simeq 0~,~~ D_2\simeq8\pi^2\frat{1}{\Delta}~.$$
For the components of tensors $E$ and $F$ similar to Eq.(\ref{t3}) we have
$$E_1\simeq\pi^2\frat{1}{\Delta}~,~~ E_2\simeq 0~,~~
F_1\simeq 2\pi^2\frat{1}{\Delta}~,~~ F_2\simeq 0~.$$
There is nothing special to get for the vector function
$$O_i=\hat\Delta_i~O~,~~O\simeq-4\pi^2\frat{\ln\Delta}{\Delta^3}~$$
and for the components of the third rank tensor $P$ similar to Eq.(\ref{t3})
$$P_1\simeq-\frat{2\pi^2}{3}\frat{1}{\Delta}~,~~
P_2\simeq-\frat{8\pi^2}{3}\frat{1}{\Delta}~.$$
The fourth rank tensor $H$ reads
\begin{eqnarray}
\label{s4}
H_{ijkl}&=&(\delta_{ij}\delta_{kl}+
\delta_{ik}\delta_{jl}+\delta_{il}\delta_{jk})~H_1+
 (\delta_{ij}\hat\Delta_k\hat\Delta_l+ \delta_{ik}\hat\Delta_j\hat\Delta_l+
\delta_{jk}\hat\Delta_i\hat\Delta_l)~H_2+\nonumber\\
&+&(\delta_{il}\hat\Delta_j\hat\Delta_k+
\delta_{kl}\hat\Delta_i\hat\Delta_j+ \delta_{jl}\hat\Delta_i\hat\Delta_k)~H_3+
\hat\Delta_i\hat\Delta_j\hat\Delta_k\hat\Delta_l~H_4~,
\nonumber
\end{eqnarray}
with the following asymptotics of its components
$$H_1\to -\frat{13}{36}\frat{\pi^2}{\Delta^3}~,
~~H_2\to -\frat{53}{12}\frat{\pi^2}{\Delta^3}~,~~H_3\to
\frat{1}{2}\frat{\pi^2}{\Delta^3}~,~~H_4\to
-\frat{5}{2}\frat{\pi^2}{\Delta^3}~.$$
For the space components of entirely symmetric tensor of the fourth rank
$Y$ similar to Eq.(\ref{t4}) we receive
$$Y_1\to \frat{1}{6}\frat{\pi^2}{\Delta}~,~~Y_2\to 0~,~~Y_3\to 0~.$$
Besides, the contribution of tensor with two time indecies $Y_{4ij4}$
(similar to Eq.(\ref{t2})) is non-zero as well. Its components
$Y^{44}_{1,2}$ in the limit $\Delta \to\infty$ are the following
$$Y^{44}_1 \to \frat{1}{6}\frat{\pi^2}{\Delta}~,~~
Y^{44}_2 \to\frat{33}{16}\frat{\pi^2}{\Delta}~.$$
Now making use of the definitions of $\Lambda_n$ (Eq.\ref{Ln}) and
$M^{ak}_n$ (Eq.\ref{Mn}) we find the interaction $S^\delta_{int}$ contributing
the terms of the second order in the parameter
$\frat{1}{g}\frat{e}{4\pi}$ to the action. Those terms are quite comparable
with the contribution of $S^d_{int}$ in the regime interesting for physics
when the source intensity $e$ is of the same order as $g$ (the factor $4\pi$
is obviously compensated while integrating). The suppression of multipole
contributions is expected to come from the geometric factor which emerges
while integrating over the azimuthal angle and for the field configurations
rather homogeneous in space, approximately it behaves
like $\frat{1}{2k+1}$ where $k$ is the multipole type. Fig.2 where the
contributions of $S^d_{int}$ (two upper curves) and $S^\delta_{int}$ (two
lower curves) are exhibited manifestly demonstrates the role of
interactions of deformable $PPs$ (solid curves correspond to the
instanton and the dashed ones correspond to the anti-instanton).

\section{Developing an approximate scheme of deformations}
Now we are well equipped to find out the optimal instanton deformations.
As mentioned above the corresponding formulae can be obtained
analytically if we neglect the
contribution of $S^\kappa_{int}$ Eq.(\ref{10b}) which is distance dependent.
Optimizing the quadratic form in the action Eq.(\ref{10}) we receive the
following system of algebraic equations
\begin{eqnarray}
\label{alsys}
&&\kappa~\delta_i\rho+\widetilde v^{al}_{ji}~ \delta_j \omega^{al}+
\Lambda_i=0~,\nonumber\\
&&\sigma^{kl}_{ij}~\delta_j\omega^{al}+\widetilde v^{ak}_{ij}~
\delta_j \rho+ \bar\eta_{mij}\varepsilon^{abc}v_2~
(2~\omega^{bm}\delta_j\omega^{ck}-\omega^{bk}\delta_j\omega^{cm})-\\
&&-v_2~\bar\eta_{kij}\varepsilon^{abc}\omega^{bn}\delta_j\omega^{cn}
-2v_1~\varepsilon_{kmn}\varepsilon^{abc}\omega^{bm}\delta_i\omega^{cn}
+\frat43 v_2~ \omega^{ak}\omega^{cn}\delta_i\omega^{cn}+M^{ak}_i=0~. \nonumber
\end{eqnarray}
In order to resolve it we introduce a new variable
$$y^{jki}=\omega^{aj}\delta_i\omega^{ak}~.$$
Recalling the orthogonality property of the rotation matrices and invariance
of the group measure Eq.(\ref{ort}) we have
\begin{eqnarray}
&&\sigma_1~ \omega^{al}y^{lki}+\sigma_2~ \delta_{ki}\omega^{al}y_{lnn}-
\sigma_2~\omega^{al}y^{lik}+2v_2~\varepsilon_{mij}\varepsilon_{dml}
~\omega^{ad}y^{lkj}- v_2~\varepsilon_{mij}\varepsilon_{dkl}
~\omega^{ad}y^{lmj}-\nonumber\\
&&-v_2~ \varepsilon_{kij}\varepsilon_{dnl}~\omega^{ad}y^{lnj} -2v_1~
\varepsilon_{kmn}\varepsilon_{dml}~\omega^{ad}y^{lni}+
\frat43 v_2 ~\omega^{ak}y^{nni}+\widetilde v^{ak}_{ij}\delta_j\rho
+M^{ak}_i=0~.\nonumber
\end{eqnarray}
Multiplying it by the  $\omega$ matrix and
calculating the products of the antisymmetric tensors we discover
the equations
\begin{eqnarray}
\label{finsys}
&&\sigma_1~ y^{ijk}-\sigma_2~ y_{ikj}+2 v_1~ y^{jik}+v_2~ y^{jki}+
v_2~ y^{kij}+\nonumber\\
&&+\delta_{jk}(\sigma_2~ y^{inn}-v_2~ y^{nin}+v_2~ y^{nni})-
v_2~\delta_{ki}(y^{jnn}+y^{nnj})+\\
&&+\delta_{ji}\left[-v_2~ y^{nkn}+v_2~ y^{knn}+\left(
\frat43 v_2-2v_1\right)~y^{nnk}\right] +v^{ij}_{kn}~\delta_n\rho
+\omega^{ai}M^{aj}_k=0~.\nonumber
\end{eqnarray}
The solution of this equation system is essentially based on the presence
of the convolution of independent variable $y_{ijk}$.
It is clear these equations are not contradictory if the following constraints
on three possible convolutions $y^{ikk}$, $y^{kik}$, $y^{kki}$ are obeyed
\begin{eqnarray}
\label{sver}
&&(\sigma_1+2\sigma_2)~y^{ikk}+(2v_1-3v_2)~y^{kik}+
\left(\frat{13}{3}v_2-2v_1\right)y^{kki}+\omega^{ai}M^{ak}_k
+v^{ik}_{kn}~\delta_n\rho=0~,\nonumber\\
&&(\sigma_2+3 v_2)~y^{ikk}-(\sigma_2+3v_2)~y^{kik}+
(\sigma_1-4v_1+4v_2)~y^{kki}+\omega^{ak}M^{ak}_i
+v^{kk}_{in}~\delta_n\rho=0~,\\
&&(\sigma_2+2v_1-v_2)~y^{ikk}+(\sigma_1-2v_2)~y^{kik}+
\left(\frat{v_2}{3}-\sigma_2-2v_1\right)y^{kki}+\omega^{ak}M^{ai}_k
+v^{ki}_{kn}~\delta_n\rho=0~,\nonumber
\end{eqnarray}
For what follows it is practical to analyze a more general situation of
the system with arbitrary constraints. The first five
terms of Eq.(\ref{finsys}) which do not contain
the convolutions might be presented using the following auxiliary tensor
$$s^{jp\beta;kq\gamma}= s_1~\delta_{jk}\delta_{pq}\delta_{\beta\gamma}+
s_2~\delta_{jk}\delta_{p\gamma}\delta_{\beta q}+
s_3~\delta_{jq}\delta_{pk}\delta_{\beta\gamma}+
s_4~\delta_{jq}\delta_{p\gamma}\delta_{\beta k}+
s_5~\delta_{j\gamma}\delta_{pk}\delta_{\beta q}~, $$
with the components $s_1=\sigma_1$, $s_2=-\sigma_2$, $s_3=2 v_1$,
$s_4=s_5=v_2$. The inverse tensor has the following form
$$\tau^{io\alpha;jp\beta}= \tau_1~\delta_{ij}\delta_{op}\delta_{\alpha\beta}+
\tau_2~\delta_{ij}\delta_{o\beta}\delta_{\alpha p}+
\tau_3~\delta_{ip}\delta_{oj}\delta_{\alpha\beta}+
\tau_4~\delta_{ip}\delta_{o\beta}\delta_{\alpha j}+
\tau_5~\delta_{i\beta}\delta_{oj}\delta_{\alpha\beta}+
\tau_6~\delta_{i\beta}\delta_{op}\delta_{\alpha j}~, $$
and it is valid for this tensor $\tau^{io\alpha;jp\beta}
s^{jp\beta;kq\gamma}= \delta_{ik}\delta_{oq}\delta_{\alpha \gamma}$
and its components are
$$\tau_1=\frat{11\cdot 199}{D}~,~~\tau_2=-\frat{25\cdot 31}{D}~,~~
\tau_3=-\frat{35\cdot 37}{D}~,$$
$$\tau_4=\tau_5=\frat{109}{D}~,~~\tau_6=\frat{7\cdot 23}{D}~,
~~D=13\cdot 83~\beta~.$$
In principle these 'strength' coefficients could be presented in the terms
of corresponding combinations of $\sigma_i$ and $v_i$. However, they look like
too  much sophisticated and it is the main reason for showing their overt
expressions here for  $PP$ in the singular gauge.
It is clear that the relations obtained are enough to get the solution of
the system Eq.(\ref{finsys}). The convolutions are extracted from the
constraint equations Eq.(\ref{sver})

\begin{center}
\vspace{0.25cm}
\parbox[b]{3.6in}{$
\left( \begin{array}{l} y^{ikk}\\
y^{kik}\\ y^{kki} \end{array}
\right)= \left\|
\begin{array}{rrr}
-\frat{136}{25~\beta}&\frat{24}{175~\beta}&\frat{8}{5~\beta}\\
&-\frat{24}{7~\beta}&\\ \frat{16}{25~\beta}
&-\frat{144}{175~\beta}&-\frat{8}{5~\beta}
\end{array}
\right\| 
\left( \begin{array}{l}
\omega^{ai}M^{ak}_k+(v_1-2v_2)\delta_i\rho\\
\omega^{ak}M^{ak}_i+3v_1\delta_i\rho\\
\omega^{ak}M^{ai}_k+(v_1+2v_2)\delta_i\rho
\end{array} \right)~.  $}
\vspace{0.25cm}
\end{center}
\noindent
When it is used in the equations for $\delta\rho$ we obtain
\begin{equation}
\label{12}
-\frat{53}{200}~\beta~\delta_n\rho-\frat{28}{25}~\omega^{ak}M^{ak}_n
+\frat{19}{25}~\omega^{an}M^{ak}_k-\frat{2}{5}~\omega^{ak}M^{an}_k
+\Lambda_n=0~.
\end{equation}
Making use of the tensor $\tau$ which is inverse to the tensor $s$
of the system Eq.(\ref{finsys}) we find out the solution $y$ and
reconstruct the solution $\delta_i \omega^{ak}=\omega^{aj}y^{jki}$.
The overt formulae are too cumbersome and are not shown here.

\begin{figure*}[!tbh]
\begin{center}
\includegraphics[width=0.75\textwidth]{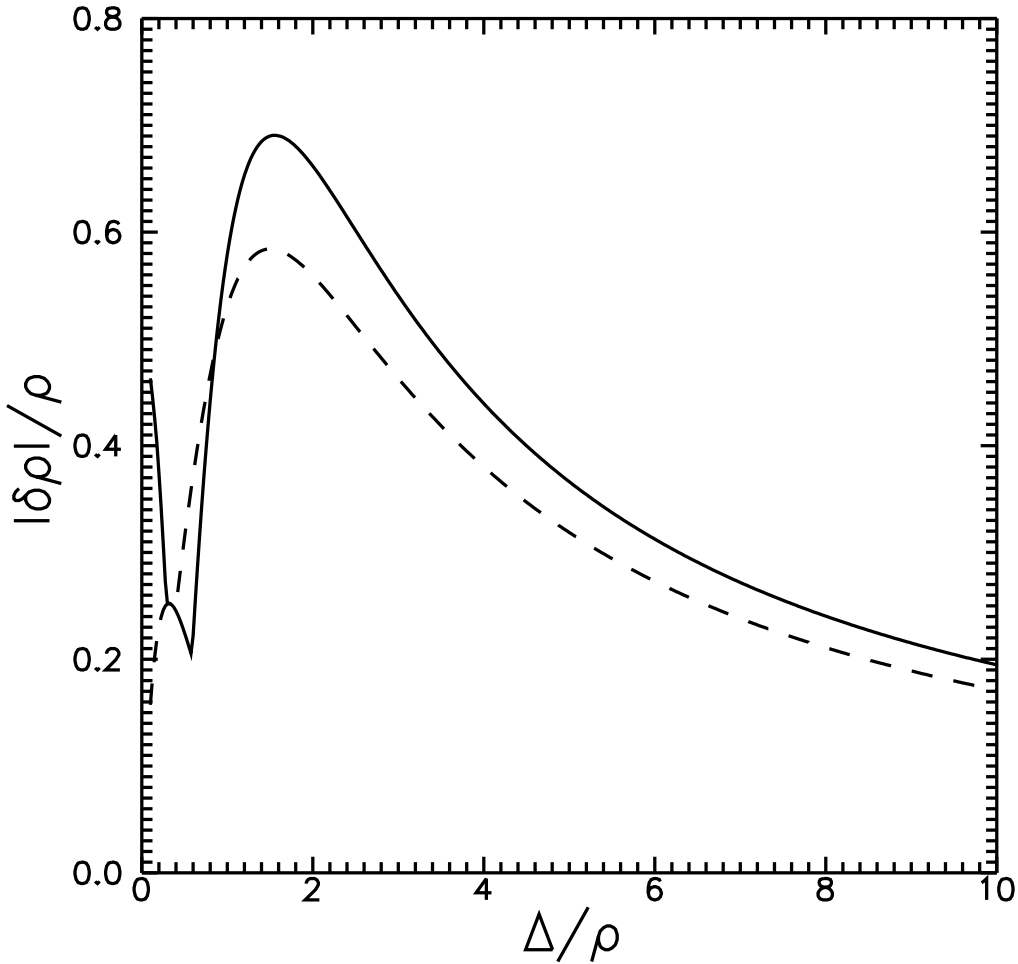}
\end{center}
\vspace{-7mm}
\caption{The norm of instanton size deviations $|\delta \rho|$ for
version with the parameters mentioned in Fig.1 caption. The dashed
line corresponds to the anti-instanton.}
\label{drho}
\end{figure*}

\section{Analysing numerical results}
The most interesting question we are going to answer here is what are
the distances where the application of the perturbative approach developed
is still valid and reliably justified. Actually, the formal criterion could be
grounded on the deformation smallness on the scale of characteristic $PP$
size which corresponds to the distance of order $|y|\simeq 1$ in the
dimensionless variables. Fig.3 illustrates the behaviour of the norm
of instanton size deviations
$$|\delta\rho|=\max_i |\delta_i\rho|~,~~i=1,2,3~,$$
as the function of the path. That is clearly seen the perturbative approach is
completely relevant at small distances and distances commensurate with several
instanton sizes. Besides, we could also conclude the deviations from
asymptotic $PP$ size are strongly dependent on the mutual positions of $PP$
and point-like source in the isotopic space. The example of Fig.3 is one
of the most rigorous limitations in $|\delta\rho|$. If one orientates an
instanton along the third axis of isotopic space
collinearly with the point-like source the deformations get considerably
smaller.  This configuration in isotopic space is characterized by the
prevailing contribution of rotational mode to $S^\delta_{int}$
Eq.(\ref{10a}) comparing to the dilatational one. However, if one orientates
$PP$ along the direction of point-like source vector, the corresponding
contribution of dilatational mode becomes prevailing, i.e. it is more
advantageous for $PP$ to rotate in the isotopic space being correlated
with the point-source. Fig.4 displays the total increase of instanton
action ($S-\beta$) (solid curve) and anti-instanton (dashed curve) as
function of the path. The upper solid line shows the contribution of
$S^d_{int}$ from Eq.(\ref{10a}) in the proximity of the coordinate origin.
Let us notice here that at large distances both the instanton
trajectory and the anti-instanton one are running above $S^d_{int}$ which
is just the action corresponding to non-deformed instanton. It means the
crumpled $PP$ develops the action maximum in this region and, hence,
the paths in the functional space might exist
and the $PP$ action might be decreasing while going along them.

In order to appoint those we consider more general
problem if determine the extremal value of the functional
Eq.(\ref{10}) with nine arbitrary constraints
$$y^{ikk}=a_i~,~~y^{kik}=b_i~,~~y^{kki}=c_i~.$$
The action Eq.(\ref{10}) with the Lagrange multipliers ${\vf \lambda}$,
${\vf \mu}$, ${\vf \nu}$ might be transformed into the form
\begin{equation}
\label{acsv}
S\to \hat S= S+\lambda_i(y^{ikk}-a_i)+\mu_i(y^{kik}-b_i) +\nu_i(y^{kki}-c_i)~.
\end{equation}
In fact, it is not easy to diagonalize directly the quadratic form for
this system. Nevertheless, the useful information on the general construction
of this function can be extracted. Trying to do that we rewrite the part
related to the rotations in the isotopic space
\begin{eqnarray}
\label{sigm}
&&S^y=y^{ip\alpha}~\sigma^{ip\alpha;jp\beta}~y^{jp\beta}~,\nonumber\\
&&\sigma^{ip\alpha;jp\beta}= \frat{\sigma_1}{2}~\delta_{ij}\delta_{pq}
\delta_{\alpha\beta} -\frat{\sigma_2}{2}~\delta_{ij}\delta_{p\beta}
\delta_{\alpha q} +v_2~ \delta_{iq}\delta_{p\beta}\delta_{\alpha j}
+v_1~ \delta_{iq}\delta_{p j}\delta_{\alpha \beta}+\\
&&+\frat{\sigma_2}{2}~\delta_{ij}\delta_{p\alpha}\delta_{q \beta}
-v_2 ~\delta_{iq}\delta_{p\alpha}\delta_{j \beta}
-v_2 ~\delta_{i\alpha}\delta_{p\beta}\delta_{j q}
-v_2 ~\delta_{i\beta}\delta_{p\alpha}\delta_{jq}
+\left(\frat23v_2-v_1\right)\delta_{ip}\delta_{\alpha\beta}\delta_{jq}~,
\nonumber
\end{eqnarray}
using the multiplet expansion in the form
$$y^{jq\beta}=y^{\{jq\beta\}}+y^{[jq\beta]}+y^{\{j[q\}\beta]}+
y^{[j\{q]\beta\}}~,$$
where the combinations symmetric in indecies are denoted by the
curly brackets and antisymmetric ones are given by the square brackets.
Then the terms without convolutions can be written in the diagonal form as
\begin{eqnarray}
\label{mult}
&&S^y=y^{ip\alpha}\left[r_1~y^{\{ip\alpha\}}+
r_2~y^{[ip\alpha]}+ r_3~y^{\{i[p\}\alpha]}+r_4~y^{[i\{p]\alpha\}}
+  \right.\nonumber\\
[-.2cm]
\\[-.25cm]
&&\left. +\delta_{p\alpha}\left(\frat{\sigma_2}{2}~y^{ikk}-v_2 ~y^{kik}
+v_2~ y^{kki}\right)-v_2~ \delta_{i\alpha} y^{kkp}+
\delta_{ip}\left(\frat23~ v_2-v_1\right)y^{kk\alpha}\right]~,\nonumber
\end{eqnarray}
where the eigenvalues are defined as
\begin{eqnarray}
\label{multcof}
&&r_1=\frat{\sigma_1}{2}-\frat{\sigma_2}{2}+v_2+v_1=\frat{13}{12}\beta~,~~
r_2=\frat{\sigma_1}{2}+\frat{\sigma_2}{2}+v_2-v_1=\frat{1}{8}
\beta~, \nonumber\\
[-.2cm] \\
[-.25cm]
&&r_3=\frat{\sigma_1}{2}+\frat{\sigma_2}{2}+v_1-v_2=\frat{11}{24}\beta~,~~
r_4=\frat{\sigma_1}{2}-\frat{\sigma_2}{2}-v_2-v_1=\frat{1}{4}\beta~. \nonumber
\end{eqnarray}
As all $r_i$ are positive we find every separate multiplet to be stable
and clearly distinguishable because of their contribution to the term of
kinetic energy type. The system of equations with arbitrary constraints
imposed looks like
\begin{eqnarray}
\label{lmn}
&&\kappa \delta_i\rho+v_1 y^{kki}+v_2 y^{kik}+v_2 y^{ikk} +\Lambda_i=0~, \\
&&\sigma_1 y^{ijk}-\sigma_2 y_{ikj}+2 v_1 y^{jik}+v_2 y^{jki}+ v_2 y^{kij}+
\delta_{jk}(\sigma_2 y^{inn}-v_2 y^{nin}+v_2 y^{nni})- v_2\delta_{ki}
(y^{jnn}+y^{nnj})
+\nonumber\\ &&+\delta_{ji}\left[-v_2 y^{nkn}+v_2 y^{knn}+
\left( \frat43 v_2-2v_1\right)y^{nnk}\right]+
\lambda_i\delta_{jk}+\mu_j\delta_{ik}+\nu_k\delta_{ij}+
v^{ij}_{kn}\delta_n\rho
+\omega^{ai}M^{aj}_k=0~.\nonumber
\end{eqnarray}
Then the conditions for this system not to be the contradictory one
might be taken as the equations for the Lagrange multipliers
\begin{eqnarray}
\label{lmnsv}
&&3{\vf \lambda}+{\vf \mu}+{\vf \nu}+ (\sigma_1+2\sigma_2){\vf a}+
(2v_1-3v_2){\vf b}+
\left(\frat{13}{3}v_2-2v_1\right){\vf c}+
\omega^{ai}M^{ak}_k +(v_1-2v_2)\delta{\vf \rho}=0~,\nonumber\\
&&{\vf \lambda}+3{\vf \mu}+{\vf \nu}+ (\sigma_2+3 v_2){\vf a}-(\sigma_2+3v_2)
{\vf b}+ (\sigma_1-4v_1 +4v_2){\vf c}+\omega^{ak}M^{ak}_i +(v_1+2v_2)
\delta{\vf \rho}=0~,\nonumber\\ &&{\vf \lambda}+{\vf \mu}+3{\vf \nu}+
(\sigma_2+2v_1-v_2){\vf a}+(\sigma_1-2v_2){\vf b}+
\left(\frat{v_2}{3}-\sigma_2-2v_1\right) {\vf c}+
\omega^{ak}M^{ai}_k +3v_1 \delta{\vf \rho}=0~.\nonumber
\end{eqnarray}
The equation for the size deformation reads
$$\kappa \delta {\vf \rho}-v_2 {\vf a}+v_2 {\vf b}-v_1 {\vf c}+
{\vf \Lambda}=0~.$$
These equations provide us with ${\vf \lambda}$, ${\vf \mu}$, ${\vf \nu}$
and putting those in Eq.(\ref{lmn}) we come (using the inverse tensor
$\tau$) to the solution for the system with constraints. At large distances
from the charge where the interaction is negligible or in the particular
case of the interaction absent (the deformed $PP$ only
is considered) the Lagrange multipliers ${\vf \lambda}$, ${\vf \mu}$,
${\vf \nu}$ and solution $y^{ijk}$ are the functions of constraints
${\vf a}$, ${\vf b}$, ${\vf c}$ only

\begin{center}
\vspace{0.25cm}
\parbox[b]{3.6in}{$
\left( \begin{array}{l}
{\vf \lambda}\\ {\vf \mu}\\ {\vf \nu} \end{array} \right)= \left\|
\begin{array}{ccc} -\frat{83}{D_1}&-\frat{43}{D_1}&-\frat{49}{3~D_1}\\
-\frat{43}{D_1}&-\frat{353}{D_1}&\frat{571}{3~D_1}\\
-\frat{49}{3~D_1}&\frat{571}{3~D_1}&-\frat{557}{9~D_1}
\end{array}
\right\| 
\left( \begin{array}{l} {\vf a}\\{\vf b}\\
{\vf c} \end{array} \right)~,$}
\vspace{0.25cm}
\end{center}
\noindent
where $D_1=\frat{5\cdot 9\cdot 32}{\beta}$.
\begin{eqnarray}
y^{ijk}&+&\delta^{jk}(A_{11}~ a_i+A_{12}~ b_i
+A_{13}~ c_i)+\delta^{ik}(A_{21}~ a_j+A_{22}~ b_j +A_{23}~ c_j)+\nonumber\\
&+&\delta^{ij}(A_{31}~ a_k+A_{32}~ b_k +A_{33}~ c_k)=0~,\nonumber
\end{eqnarray}
moreover, the matrix $A$ is just a projector necessary to excrete from the
solution $y$ nine constraints imposed $A_{11}=A_{22}=A_{33}=-\frat{2}{5}$,
$A_{12}=A_{13}=A_{21}=A_{23}=A_{31}=A_{32}=\frat{1}{10}$. Then as the result
the action Eq.(\ref{acsv}) becomes the quadratic form of three constraint
vectors ${\vf a}$, ${\vf b}$, ${\vf c}$ i.e.
$$\hat S=-\frat12 ({\vf \lambda}{\vf a}+
{\vf \mu}{\vf b}+{\vf \nu}{\vf c})~.$$
Two eigenvalues of this form are positive and one is negative.
The direction where going along the action is getting smaller is given by
the following equation
$${\vf a}={\vf 0}~,~~3\cdot 571~{\vf b}-557~{\vf c}={\vf 0}~,$$
(let us remind once more that these result was obtained considering the instanton
in singular gauge, and we omitted their representations while the
corresponding components  of tensors $\sigma_i$, $v_i$ because of their very
cumbersome forms). We did not manage to diogonalyze equations
(\ref{finsys}) manifestly nevertheless imposing arbitrary constaints
we demonstrate the part of eigenvalues of quadratic form is negative
and $PP$ can manifest itself as a 'sphaleron'. In principle studying
the deformed $PPs$ out of the perturbation theory developed here
could allow to understand transition of nonperturbative field configurations
into the perturbative ones. Usually such a transition is associated
with instanton anti-instanton annihilation only and justifies
the criticism of simple superposition ansatz in the IL model \cite{51}.
From this view point we demonstrate that already in one particle sector
there exist the ways in the functional space which inter-relate
the solutions from different topological classes.
However, it is a well known fact that the solution with nontrivial topology
in spite of the loss of action is characterised by large statistical weight.

Dealing with one $PP$ we could study the deformations which leave the
solution within the  same topological class if the condition of topological
charge conservation $\delta N=0$ is imposed. In the approximation of
quadratic deviations we receive the following result
\begin{equation}
\label{deln}
\delta N=\frat{1}{4\beta}\int d x~G^{a}_{\mu\nu}\widetilde G^{a}_{\mu\nu}-1=
-\frat{3}{10}~\delta_\mu\rho\delta_\mu\rho
-\frat{1}{12}~\delta_\mu \omega^{ak}\delta_\mu \omega^{ak}~,
\end{equation}
where $\widetilde G^{a}_{\mu\nu}=
\frat{1}{2}\varepsilon_{\mu\nu\alpha\beta}G^{a}_{\alpha\beta}$.
It leads to the conclusion that the dipole-like deformations (the expansion
Eq.(\ref{6}) is meant) are forbidden if the toplogical charge is not changed.

As an example of the system with the constraints imposed we
consider the substitution with four non-zero components
(which is dictated by the particular form of matrix $\omega$)
$$\delta_1\rho=x~,~~\delta_1 \omega^{13}=\delta_1\omega^{22}=y~,~~
\delta_1\omega^{31}=z~.$$
Besides, the contribution of terms $S^\kappa_{int}$ could be analysed
in this example. The solution is determined by minimizing the following
quadratic form
$$S-\beta=\frat{A}{2}~x^2+\frat{B}{2}~y^2+\frat{9}{16}\beta~z^2
+Fyz +Cxy+Dxz+\Lambda_1 x+E y+G z~,$$
where
\begin{eqnarray}
\frat{A}{2}&=&\frat{\kappa}{2}\mp\frat1g\frat{e}{4\pi}\Sigma_{111}+
\left(\frat{e}{4\pi}\right)^2(2Q_{11}-R_{2211}-R_{3311})~,\nonumber\\
\frat{B}{2}&=&\frat{17}{24}\beta\pm\frat{1}{g}\frat{e}{4\pi}~
(2T_{111}-S_{133}-S_{122})+ \left(\frat{e}{4\pi}\right)^2
(2U_{11}-V_{2211}-V_{3311})\pm\nonumber\\
&\pm&\frat{1}{g}\frat{e}{4\pi}~2~(Z_1-Z_2-Z_3+Z_4)~,\nonumber\\
C&=&\frat{7}{12}\beta \pm\frat{1}{g}\frat{e}{4\pi}~
2~(2W_{111}-3X_{221}-X_{331})
\pm\frat{1}{g}\frat{e}{4\pi}~4~(\Xi_1-\Xi_2-\Xi_3+\Xi_4)~,\nonumber\\
D&=&-\frat{7}{24}\beta\mp\frat{1}{g}\frat{e}{4\pi}\Psi_{111}
\pm\frat{1}{g}\frat{e}{4\pi}~2~(\Xi_1-\Xi_2-\Xi_3+\Xi_4)\pm\nonumber\\
&\pm&\frat{1}{g}\frat{e}{4\pi}~2~(3\Theta^{44}_1+\Theta^{44}_2
+21\Theta_1+12\Theta_2+\Theta_3)~,\nonumber\\
F&=&\frat{13}{24}\beta\pm\frat{1}{g}\frat{e}{4\pi}~2~(3\Phi^{44}_1+\Phi^{44}_2
+21\Phi_1+12\Phi_2+\Phi_3)\nonumber\\
E&=&M^{13}_1+M^{22}_1~,~~~~G=M^{31}_1~,\nonumber
\end{eqnarray}
here the upper sign corresponds to the instanton. The instanton trajectory
is depicted by the dotted line in Fig.4.

\begin{figure*}[!tbh]
\begin{center} \includegraphics[width=0.75\textwidth]{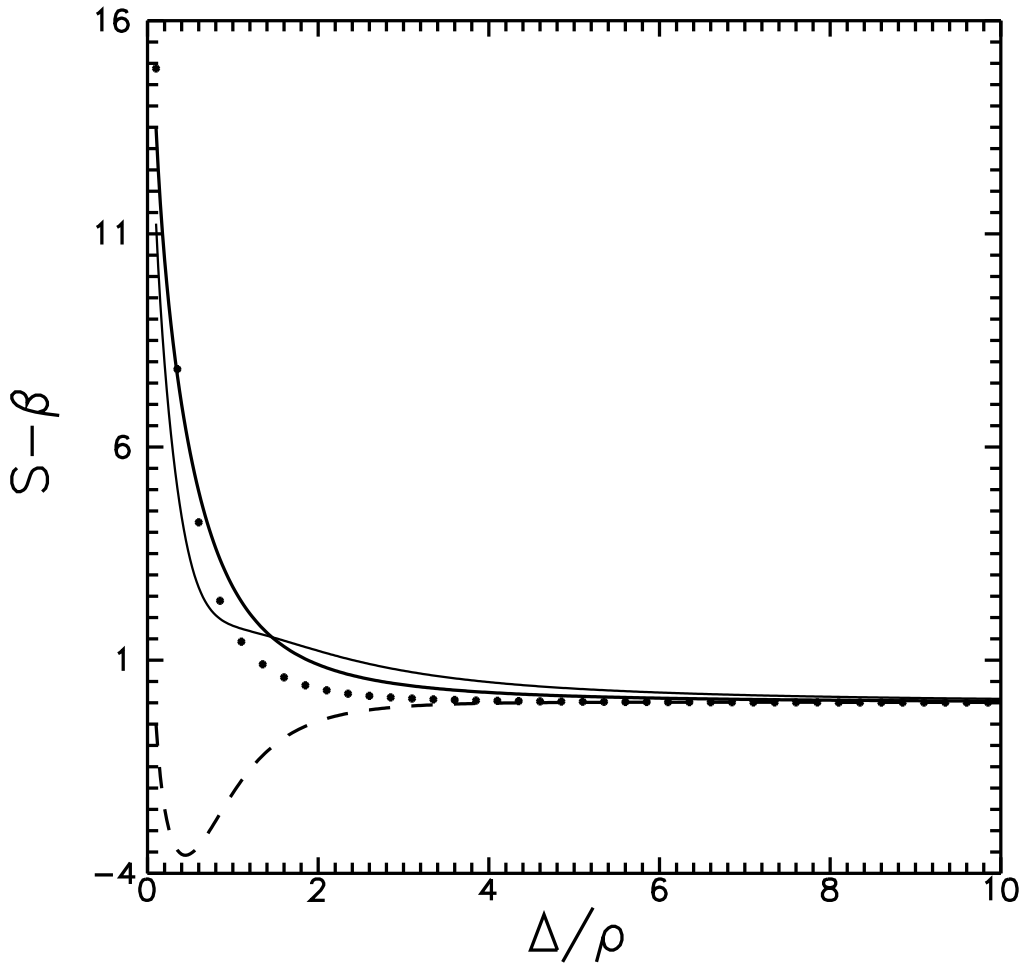} \end{center}
\vspace{-7mm}
\caption{The increase of action ($S-\beta$) for the crumpled instanton  and
anti-instanton (dashed line). The parameters of the rotation matrix and the
path are mentioned in the caption of Fig.1. Two solid lines correspond to the
instanton and the upper one (around the coordinate origin) shows $S_{int}^d$.
The dotted line shows the instanton trajectory with the constraints imposed.}
\label{stot}
\end{figure*}

The topological charge change for the trajectories under consideration is
plotted in Fig.5 and allows one to conclude that the interaction does not
take the solution out of the topological class with the charge $|N|=1$.
The quantity defined by Eq.(\ref{deln}) characterizes the deformation extent
of aproximate instanton solution only, also one should pay interest to
topological charge of superposition solution on the whole.

Another interesting remark as to the possible constraint conditions comes
from their optimization, i.e. from the analysis
of nine constraints obtained for the action extremum available. It turns out
the initial trajectories of Eq. (\ref{sver}) are just the case and the
Lagrange multipliers for this extremal action are trivial
(${\vf \lambda}={\vf \mu}={\vf \nu}={\vf 0}$).

\section{Analyzing the IL model}
The results obtained above allow one to conclude the perturbative approach
developed is applicable to effectively explore  the external field impact on
$PP$ no matter how large (or small) the distances are. One should expect the
substantial suppression of instantons at the distances of average
instanton size order in IL but the Coulomb-like
interaction extends  to the distances much larger, of several
characteristic instanton sizes.

Apparently, the configurations of crumpled instantons may come about very
essential for the IL model. The entropy of crumpled pseudoparticles and their
contribution to the functional integral are of the same order as for the
habitual instantons. (The entropy of a field configuration
is implied as the $log$ of the volume in functional space occupied by similar
configurations.) An importance of searching the similar trajectories which
have the large entropy (even losing in the action) at calculating the
functional integral is well known (see, for example,
\cite{dp}). In fact, the condition $\delta S=0$ for searching
the optimal crumpled configurations operates with the interaction
contributions coming from exponential factor only and, as can be
shown, is a valley equation in quasi-clasical approximation. In order to
pretend to higher precision and to take into account so-called quantum
interactions one should deal with the valley method (see for example
\cite{91} where the similar practical method is given).
However, the discussed example of dipole-like deformations Eq.(7)
demonstrates obviously how sophisticated this configuration can happen to
be even in this comparatively simple case.

Another remark worth mentioning here concerns the behaviour of the
pseudo-particle deformations at large distances.
Studying the asymptotic behaviour of
interaction terms one easily finds out  their Coulomb-like character,
i.e. even in far distant area the pseudo-particle field may be
noticeably polarized. In principle, if it turns out possible in some way to
detach the long-range component on the background of inter-instanton
interactions (for example, for dilute gas of very cooled (anti-)instanton
configurations) the chance may appear to measure the coupling constant $g$
on the lattice directly if an external field of intensity $e$
(which is controlled) is applied to its boundary. The dilatational and
rotational deformations may contribute to the observables only as the
corresponding averages $\langle\delta_n\rho\rangle$, $\langle
\delta\omega^{ab}_i\rangle$ which are
defined by two independent parameters $e^2$ and $\frat{e}{g}$.
It looks feasible to construct such a linear combination of those two
averages which depends on the intensity $e$ only and certainly should be
under control during the measuring process whereas
another could provide us with the magnitude of $g$. The great experience
collected in lattice QCD to work with the boundary conditions imposed
\cite{bound} gives us the hope to succeed on this way applying the
cooling technics \cite{lat} which isolates the instanton
configurations successfully and incidently they look quite deformed \cite{neg}.

\subsection{Crumpling instantons in the IL model}
Here we are trying to suggest a fairly simple approach of taking into account
the crumpled pseudoparticle configurations in the functional integral.
If we pretend to make this estimate calculating precisely the exponential
factor of generating functional only then their contribution could be found in
quasi-classical approximation just in the same way as it
is done for the undeformed pseudo-particles (see the designations in \cite{8})
\begin{equation}
\label{genfun}
Z= \sum_{N}\frat{1}{N!}~\prod_{i=1}^{N}~\int
\frat{d\gamma_i}{V~\rho_{i}^5}~C_{N_c}~
\widetilde\beta^{2N_c}~e^{-S}~,~~~~d\gamma_i=dz_i~d\rho_i~d\omega_i.
\end{equation}
As supposed in the IL model this integral is saturated by the
configurations of characteristic size $\bar\rho$. In the approach
under consideration $S$ means an action including the deformations in the
above formula. Thus, adding the contibution of the crumpled pseudo-particles
of the similar characteristic scale we are able to work
within the same precision studying the IL reaction on the external impact.
The variational Ritz method as defined by Eq.(\ref{5}) looks to be pretty hard
and incomplete from the viewpoint of the integration in the functional space.
We have permitted the velocities and higher coefficients of multipole
expansion only to vary there but not the major
expansion term that is the average instanton size. In view of the
functional integration expanding the average pseudo-particle size varied
seems more natural. Then we have for this
particular expansion the following expression
\begin{equation}
R_{in}(x,z)=\rho(z)+\frat{\partial
\rho(z)}{\partial z_\mu}~y_\mu+\frat{\partial^2\rho(z)}
{\partial z_\mu\partial z_\nu}~\frat{y_\mu~ y_\nu}{2}+\dots~,
~~~~~|y|\leq L~~.
\end{equation}
and the similar one for $\Omega$. The generating functional becomes now
\begin{equation}
\label{ILgenfun}
Z= \int D[\rho (z)]~D[\omega (z)] \sum_{N}\frat{1}{N!}~\prod_{i=1}^{N}~\int
\frat{d z_i~d\omega_i}{V}~C_{N_c}~ \widetilde\beta^{2N_c}~e^{-S}~.
\end{equation}
Remembering that the IL approach, as formulated, takes into account
interactions between the pseudo-particles we should expect in Eq.(\ref{8})
the pseudo-particle as the field $B$, too. In order to simplify the
calculations let us suppose for the moment the (anti-)instantons undergo
the insignificant changes of size because of an impact of external source
on IL, i.e. $\rho (z)=\bar\rho+\delta\rho(z)$ and colour instanton
orientation gets $\omega_i (z)=\omega_i+\delta\omega(z)$ and for
anti-instanton it is $\bar\omega_i (z)=\bar\omega_i+\delta \bar\omega(z)$.
The first terms of $\omega$ correspond to the anticipated fixed directions
in colour space. The field strength of single instanton looks like
$$G_{\mu\nu}^a(A)=-\frat{4}{g}~\omega^{ak} (z) M_{\mu\alpha} M_{\nu\beta}~
\bar\eta_{k\alpha\beta}~\frat{\rho^2 (z)}{(y^2+\rho^2(z))^2}~,$$
with $\omega$ changed for $\bar\omega$ for the anti-instanton.
When we have the instanton-instanton 'molecule' the mixed component of field
strength $G(A,B)$ reads
$$G_{\mu\nu}^a(A,B)=\frat{4}{g}~\varepsilon^{abc}\omega^{bn} (z_1) \omega^{ck}
(z_2)~(\bar\eta_{n\mu\gamma}\bar\eta_{k\nu\alpha}-
\bar\eta_{n\nu\gamma}\bar\eta_{k\mu\alpha})~\frat{y_{1\alpha}y_{2\gamma}}
{y_1^{2}y_2^{2}}~\frat{\rho^2 (z_1)}{(y_1^{2}+\rho^2(z_1))}\frat{\rho^2 (z_2)}
{(y_2^{2}+\rho^2(z_2))}~$$
where $y_i=x-z_i$. It is clear all the combinations of colour matrices for the
instanton-instanton interaction owing to the property Eq.(\ref{ort}) give the
unit matrix if we limit ourselves with the contact type of interaction only.
This approximation seems well justified since the integrals of our interest
are quickly decreasing for the far distant pseudo-particles
$|z_1-z_2|\gg\bar\rho$. For the instanton-anti-instanton
molecule such a component is given by
$$G_{\mu\nu}^a(A,B)=\frat{4}{g}~\varepsilon^{abc}\bar\omega^{bn}
(z_1) \omega^{ck} (z_2) ~ (\eta_{n\mu\gamma}\bar\eta_{k\nu\alpha}-
\bar\eta_{k\mu\alpha}\eta_{n\mu\gamma})~\frat{y_{1\alpha}y_{2\gamma}}
{y_1^{2}y_2^{2}} ~\frat{\rho^2 (z_1)}{(y_1^{2}+\rho^2(z_1))}\frat{\rho^2
 (z_2)}{(y_2^{2}+\rho^2(z_2))}~.$$
The properties of the 't Hooft symbol make it possible to demonstrate that
the product of matrices $\omega$ and $\bar\omega$ gives the unit one in this
case as well. Then the averaged integral for ineracting pseudo-particles
may be written in the following form (see Ref. \cite{8} for designations)
\begin{eqnarray}
\langle S(A,B)\rangle&=& \int \frat{d z_1}{V}~\int \frat{d z_2}{V}~\bar
U_{int}(A,B)~, \nonumber\\
[-.2cm]
\\[-.25cm]
\bar U_{int}(A,B)&=&\frat{8\pi^2}{g^2}~\frat{N_c}{N_c^{2}-1}~\int dx~
\frat{(7~y_1^{2}y_2^{2}-(y_1y_2)^2)~\rho^2(z_1)\rho^2(z_2)}
{y_1^{4}(y_1^{2}+\rho^2(z_1))^2~y_2^{4}(y_2^{2}+\rho^2(z_2))^2}~.\nonumber
\end{eqnarray}
And integrating over the positions of single detached pseudo-particle one may
receive an estimate in the form of contact term
$$\int d z_1~\bar U_{int}(A,B) \simeq \frat{8\pi^2}{g^2}~\xi^2~\rho^2(z_2)
\rho^2(z_2)\simeq \frat{8\pi^2}{g^2}~\xi^2~\bar\rho^2\rho^2(z_2)~, $$
where $\xi^2=\frat{27\pi^2}{4}\frat{N_c}{N_c^{2}-1}$.
(Obviously, the interaction integral is independent of the colour orientation
because of the IL isotropy in colour space.) Besides, we should take into
account the term of kinetic energy type Eq.(\ref{9}) for each
pseudo-particle. As the result, it has been shown in \cite{6},
the generating functional (\ref{ILgenfun}) for such a sort
of crumpled pseudo-particles takes the form of the effective Lagrangian
for the scalar field $\rho(z)$ with the mass gap
$M^2=\frat{2~(11N_c-2N_f-12)}{3~\kappa~\bar\rho^2}$.
The dependence on the rotational component ($\delta\omega$,
$\delta\bar\omega$) occurs to be trivial. The estimate for kinetic coefficient
$\kappa \sim 1.5 \beta$--- $6\beta$ has been obtained before and depends
on the ansatz for the saturating configuration.
For example, for pure gluodynamics with the parameters $N_c=3$,
$\beta \simeq 17.5$, $\bar \rho \Lambda\simeq0.37$ at $\kappa=4\beta$ it is
$M\simeq 1.21 \Lambda$. In this case we have the soft mass scale with
characteristic for IL $\Lambda$~--$280$ MeV.
The result $\kappa=0.9\beta$ we have
received in this paper corresponds to essentially
harder mass scale, above $1$ GeV. The IL excitation discussed in its
quantum numbers could be identified as a glueball.

\begin{figure*}[!tbh]
\begin{center} \includegraphics[width=0.75\textwidth]{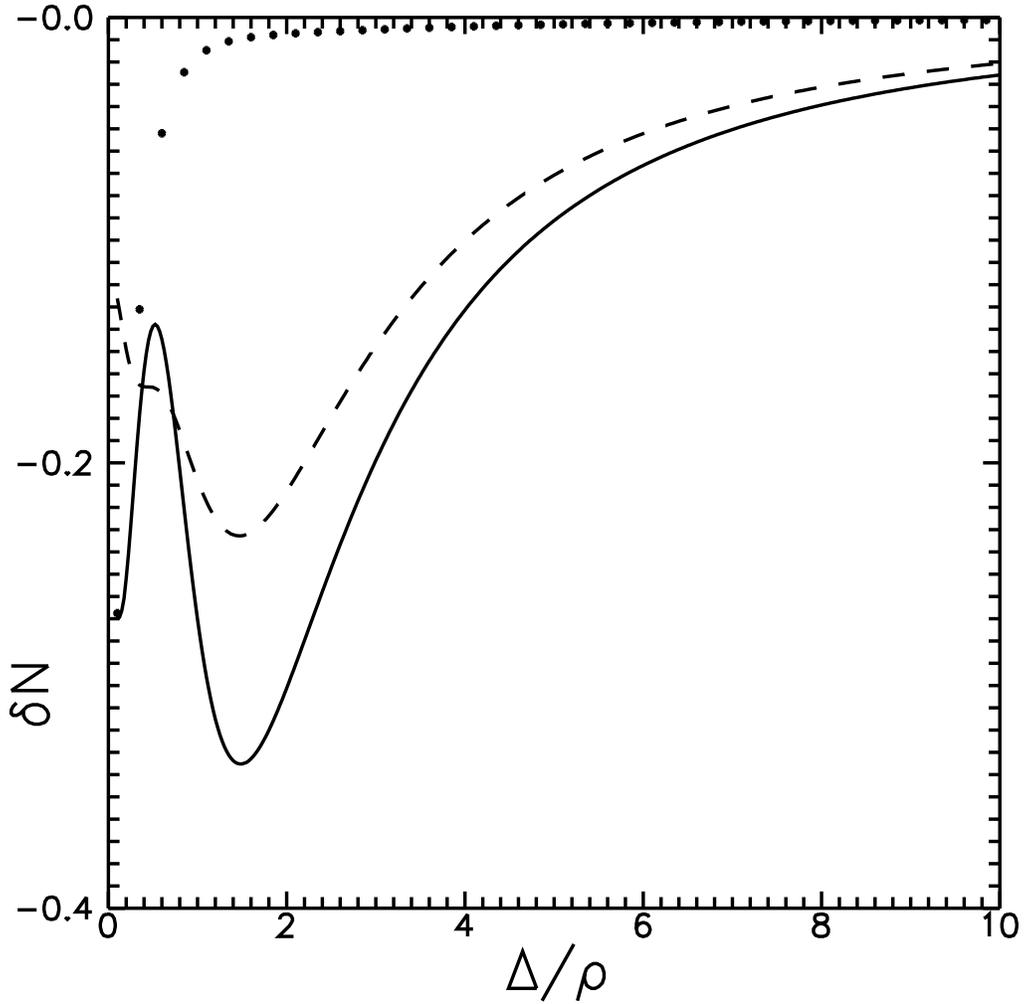} \end{center}
\vspace{-7mm}
\caption{Behaviour of the topological charge $\delta N$ for the crumpled
instanton and anti-instanton (dashed line). The parameters of rotation
matrix and path are the same as in Fig.1. The dotted line corresponds
the trajectory for instanton with the constraints imposed, see the text.}
\label{topz}
\end{figure*}

\subsection{Average source energy in IL}
In this paper we are studying the way in which an external field acts
upon the pseudo-particle. If we don't fix the features of field $B^a_{\mu}(x)$
in advance the feedback of IL could be described self- consistently. Then the
multipole expansion we used allows us to factorize the problem in a sense and
construct the closed equation for the Fourier component of the field
$B^a_{\mu}(p)$. That equation looks rather promising to calculate the Debay
screening radius that is one of the most important physical parameters
characterizing IL. However, even the description we have already elaborated
for the (anti-)instanton behaviour in
external field is pretty rich to provide us with interesting information
on the IL properties and possibility to explore the various observables.
We mention here one result only calculating an average energy of Euclidean
source inserted into IL. As a relevant saturating configuration in the IL
model one takes the superposition of (anti-)instanton fields with the
source field $B^a_{\mu}$added as
\begin{equation}
\label{sup2}
A^{a}_\mu(x)=B^a_{\mu}(x)+\sum_{i=1}^N A^{a}_\mu(x;\gamma_i)~,
\end{equation}
where $\gamma_i=(\rho_i,z_i,\omega_i)$ denotes the parameters describing
the $i$-th instanton. It is seen in Fig. 4 that at the distances larger than
$2 \Delta/\rho$ the IL density becomes practically equal to its asymptotic
value $n(\Delta)\sim n_0 e^{\beta-S}\simeq n_0$ because the action of each
pseudo-particle there approximately coincides with $\beta$,
$S-\beta\ge S^d_{int}$.  The quantity here we are interested in should be
given by averaging $S$ over the pseudo-particle positions and their colour
orientations (taking all the pseudo-particles of the same size for
simplicity) as
\begin{eqnarray}
\label{s}
&&\langle S\rangle=\prod_{i=1}^N \int
\frat{d z_i}{V}~ \int d\omega_i~~S=\nonumber\\
[-.2cm]
\\[-.25cm]
 &&=\frat{e^2}{4\pi}\frat{1}{a}~X_4+N~\beta+N \int \frat{d {\vf \Delta}}{L^3}
 \left(\frat{e}{4\pi}\right)^2\left(J+\frat{K_{ii}}{3}\right)~,\nonumber
\end{eqnarray}
where $L$ is a formal upper integration limit, $V=L^3X_4$ defines the
IL volume, $X_4$ is an upper bound of the 'time' integration, $N$ denotes
the total number of pseudo-particles and $a$ is a source size value (on strong
interaction scale, of course). Taking the asymptotic values of $J$ and $K$
functions and returning to the dimensional variables for the moment one
instanton contribution to the average action can be written
down in the following form
$$\langle S\rangle\simeq \frat{e^2}{4\pi}\frat{1}{a}~X_4+
\frat{N}{V}~\beta~L^3 X_4+\frat{N}{V} \frat{6 \pi^3}{\beta}
\frat{e^2}{g^2}~\bar\rho^2~L~X_4~, $$
where $\bar\rho$ is the mean size of (anti-)instanton in IL.
The result is given in the form with the common factor $X_4$ extracted because
our concern now is the (anti-)instanton behaviour in the source field
background where the field is steady and the solution possesses the scaling
property at any $x_4$ slice. In the limit $N,V\to\infty$ and at the IL density
$n=N/V$ fixed this result becomes
$$\langle S\rangle\simeq E~X_4~,~~E=\frat{e^2}{4\pi}\frat{1}{a}+
n~\beta~L^3 +n~\frat{6 \pi^3}{\beta} \frat{e^2}{g^2}~\bar\rho^2~L~.$$
Formally, the last term of $E$ looks like a small correction to the gluon
condensate (the second term). However, this contribution linearly increasing
with $L$ and proportional to $e^2$ has a different physical meaning of an
additional contribution to the source self-energy
$$E \simeq \sigma~L~,~~~\sigma=n~\frat{6 \pi^3}{\beta}
\frat{e^2}{g^2}~\bar\rho^2~.$$
The tension value $\sigma$ for the IL characteristic parameters
$\frat{\bar\rho}{\bar R}\simeq \frat{1}{3}$
where where $\bar R$ is the mean separation of pseudo-particles,
$n=\bar R^{-4}$, $\beta\simeq 12$, $\bar\rho\simeq 1$ $GeV^{-1}$
\cite{2},\cite{3} comes about $\sigma \simeq 0.6$ GeV/fm
(if one takes for the brief estimate for the source intensity $e\simeq g$).
This result looks fully relevant in view of the qualitative character of
estimates which IL is able to provide.  Moreover, such a value is in
reasonable agreement with the estimates extracted from the potential models
for heavy quarkonia, for example. If one intends to explore the magnitudes
like $\langle S_{int}~e^{-S_{int}}\rangle$
(which could model an effect of suppressing the pseudo-particle
contribution in the source vicinity) in numerical calculations it becomes
evident the linearly increasing behaviour starts to form at
$\Delta/\bar\rho\sim 3$---$4$.  An interesting phenomenon
of energy increase with a distance increasing which we discovered above
looks promising for the applications in IL. Indeed, this result teaches the
source mass (we should treat an additional contribution found out just in
this way) is unboundedly increasing (if the screening mechanism is not
included) what could be interpreted as preventing the
possibility to immerse a bare colour charge into IL. Our estimate of
asymptotic energy of the Euclidean source put into IL shows the major
contribution to the generating functional in quasi- classical approximation
while all coupling constants are fixed at the scale of average instanton size
$\bar\rho$.

Further we are going to exploit this interesting feature of IL
for analysis of the colour dipole behaviour dealing with the point-like
sources $e\delta^{a3}$, $-e\delta^{a3}$ which are oriented along the third
axis of isotopic space and placed in the points ${\vf z}_1$ and  ${\vf z}_2$,
respectively. Then the strength tensor component
of our interest looks as follows
$$G_{4i}^a(A,B)=2~ \frat{e}{4\pi}~\varepsilon^{a3c}~
\Omega^{ck}~\bar\eta_{ki\alpha}~
\frat{y_\alpha}{y^2}~\frat{\rho^2}{(y^2+\rho^2)}~
 \left(\frat{1}{|{\vf y}+{\vf \Delta}_1|}-\frat{1}{|{\vf y}+
{\vf \Delta}_2|}\right)~,$$
where ${\vf \Delta}_{1,2}={\vf z}- {\vf z}_{1,2}$. The contribution to the
average energy $E$ at large distances which is additional to the Coulomb one
is defined by the integral similar to that for the configuration with
single source (the same coefficient but
distinguishing average over the instanon positions), i.e.
$$I_d=\int d{\vf \Delta}_1 \left(\frat{1}
{{\vf\Delta}_1^{2}} -2~\frat{1}{|{\vf\Delta}_1|
|{\vf\Delta}_2|} +\frat{1}{{\vf\Delta}_2^{2}}\right)~.$$
When the distance between the sources $l=|{\vf\Delta}_1-{\vf\Delta}_2|$ is
going to zero the field disappears and the final result should be zero.
The integral may be dependent on two parameters $L$ and $l$ only.
The dimensional analysis reveals the integral
is a linear function of both parameters but the $l$-dependence only obeys
the requirement of integral petering out at $l\to 0$. In order to determine
the coefficient we find easily
$$\frat{I_d}{4\pi}=L-2\left(L-\frat{l}{2}\right)+L=l~,$$
(here the contributions of three integrals are shown separately),
where $l=|{\vf\Delta}_1-{\vf\Delta}_2|$ is the distance between charges.
And finally we have for the additional contribution to the dipole average
energy while in IL $$E\sim \sigma l~.$$ The non-singlet in colour states
are strongly suppressed in IL by the factor of order of the screening radius
$\sim\sigma~R_D$.

Finally, we would like to make one general remark which is
not directly related to IL but seems to be of heuristical importance.
We encounter the crumpled configurations of topological nontrivial
solutions (pseudo-particles) not only while studying the impact of external
fields on them or analyzing their interactions.
We discover such configurations, for example, studying the
simplest kinks in the problem of double well potential
$V(x)=-\lambda(x^2-\eta^2)^2$ (we are dealing here with an analysis in the
Euclidean space). Let us consider the kink at finite interval $(-T,T)$
where $T\gg \omega^{-1}$ if the latter is a characteristic kink size
($\omega^2=8\lambda\eta^2$). The approximate solution is given by
$x_k(t)\simeq x_{as}(t)=\eta~{\mbox{th}} \frat{\omega t}{2}$ and at the turning points
$x_k(T)=x_k(-T)\simeq \eta$ the particle velocity
disappears $\dot x_k(\pm T)=0$. The trajectory choice is dictated by the
requirement that a particle should pass through the coordinate origin at
zero time moment (the coordinate of kink center $t_c=0$). Apart from this
(unique) symmetric kink the trajectories
with the biased center which crosses the axis $x=0$ not only at zero time
moment are of special interest. In the vicinity of kink center such
solutions are approximately described by $x_{c}(t)=\eta~{\mbox{th}}
\frat{\omega (t-t_c)}{2}$. Then, in order to
construct these solutions we consider the time interval larger than the
initial one $(-T_2,T_2)$ where $T_2>T$. There is the trajectory with the
kink center located at the coordinate origin which we designate as
$\acute x_k(t)$, i.e. $\acute x_k(t)\simeq x_{as}(t)$, and this particle
starts and finishes again with the zero velocity $\dot x_k^{'}(\pm T_2)=0$.
Calculating the energy of this particle we find the magnitude
which is slightly larger than for the trajectory $x_k(t)$ and the particle
climbing the saddle point reaches the points higher
than the turning points $x_k(\pm T)$ of
the initial trajectory. Now, let us shift the time interval origin for the
trajectory $\acute x_k(t)$. Thus, we receive the solution
$x_c(t)=\acute x_k(t+T-T_2)$ with the baised center. For this
configuration the particle starts to fall down at the time moment $-T$
to the well with the zero initial velocity and crosses over the axis
$x=0$ at the time moment $T_2-T$. Then climbing the top again it does
not stop at the time moment $T$ and continues to move with a
certain velocity. Hence, we need to install a hard barrier to reflect
it compelling to move in the opposite direction. Strictly speaking such
trajectories are not the equation solutions for the initial potential.
The trajectory of particle finishing 'properly' could be constructed by a
 similar way, however, in distinction from the kink
$x_k(t)$ this trajectory $x_c(t)$ is asymmetric. Its right hand wing does
not copy absolutely the left hand one. Truely, the difference
is quite inconsiderable (exponentially small) if $t_c\ll T$
but becomes very essential when the shift of kink center is of the
same order as the interval magnitude $T$.
Generally, the particle could have non-zero velocity both at a start
and at a finish. If we remember now that calculating the functional integral
we are limited by the special requirement for trajectories to deal
with only those which have both ends fixed. Our discussion above teaches it
could be satisfied by breaking the equation of motion.

Thus, we are forced to examine the inconventional kinks which are just
the particular set of trajectories with the measure in the functional space
proportional to the magnitude of interval $\int d t_c =2~T$.
The kink-anti-kink configurations form a much more sophisticated set
of trajectories (with complicated interaction between fragments) which are
transformed into standard oscillator (perturbative) trajectories,
with oscillation frequency $\Omega^2=4~\lambda~\eta^2$,
while approaching the well bottom. Such trajectory should not be
identified with the kink-anti-kink configurations.
The instructive message of this example, at least, in the context of previous
study is that the exact definition of a kink is of less importance compared
to the necessity of taking into account properly the contribution of adequate
trajectories in the functional integral. We mean the trajectories of
considerable power in the functional space, for example, integrating over
their deformations if those contribute significantly to an effective action.

\section*{Conclusion}
Since the instanton discovery a suggestive way to access an information
on the nonperturbative QCD dynamics is to investigate the role of this
coherent gluon fields as major objects populating the QCD vacuum.
The development of the IL model allowed to obtain the estimates, the bulk
instanton properties in vacuum and the lattice simulations have later
supported these. Hovewer, the study of instanton behaviour in the
external fields, as was understood, is of great importance for resolving
the principle QCD problems but the results here are still rather hazy.
It was a main motivation to have in this paper as a focal point
analysis of the $PP$ behaviour in the field of point-like source.
In order to deal with such a task we developed the perturbation
approach based on the variation of the instanton parameters. We have
formulated the variational problem of searching  the optimal $PP$
deformations which is algebraically resolved by the Ritz method 
within the multipole expansion. The regions of parameter changes 
interesting for the applications in the IL model were especially 
analyzed in detail. We demonstrate the solution of algebraic 
problem might be essentially simplified  and  the overt formula
for the deformation fields in dipole approximation might be obtained 
if one neglects the contribution to the term of kinetic energy type 
$S^\kappa$ which depends on the distance to the perturbation source. 
The coefficients of the quadratic form in this term of kinetic 
energy type $S^\kappa$ are related to $PP$ itself and
charaterize, in a sense, its 'pliablity' for changing 
the size and orientation in the 'isotopic' space. The terms responsible 
for the interactions between undeformed $PP$ and $PP$ with unchanged 
parameters with fixed point-like Euclidean source of colour
field were calculated. In addition, it was realized the dipole-like 
contributions to $S_{int}^{d}$ which are proportional to $e/g$ occur 
noticeably less in the parameter region
characteristic for the IL model at smaller distances than the terms of
$e^2$-order. The effective and rather simple method to take into 
account the 'crumpled' configurations at calculating approximately 
the functional integral. The description of instanton behaviour in 
the field of point-like Euclidean colour source developed in the
paper clearly demonstrates that the valley topography in the 
functional space might be so sophisticated that makes it impossible 
to investigate the problem in any general formulation. Finally, in 
the framework of superposition ansatz the estimate of
average energy of non-abelian dipole in the IL medium has been 
found out. This energy for the dipole in colour singlet state 
escalates linearly with the separation increasing for the 
point-like sources unscreened. The corresponding
coefficient of this dependence develops the 
magnitude similar numerically to one as
inferred from the lattice OCD calculations.

\noindent Our work was partially supported by the Grants STCU \#P015c,
CERN-INTAS 2000-349, NATO 2000-PST.CLG 977482.

\section*{Appendix}
Here we present the integrable terms. We show only the components of the
corresponding tensors which are mentioned in the paper The interacting
terms linear in $\delta \rho$ are the following
\begin{eqnarray}
&&A_1=\frat{4~\pi^2}{\Delta^2}\left[\frat83~D^5-\frat83~\Delta^5
-\frat{20}{3}~D^3+\frat92~D-\frat{3}{D}+\frat{5}{2\Delta}~
\ln(\Delta+D)\right]~, \nonumber\\
 &&A_2=\frat{4~\pi^2}{\Delta^2}\left[4~\Delta^5-4~D^5
+10~D^3-\frat{7}{2}~D+\frat{3}{D}
+\frat{2}{D^3} -\frat{15}{2\Delta}~\ln(\Delta+D)\right]~,\nonumber
\end{eqnarray}
where $\hat\Delta_i=\frat{\Delta_i}{\Delta}$ is a unit tensor. The term $jA$
of action Eq.(\ref{1}) contributes to $A_2$.  The interacting terms linear in
$\delta \Omega$ take the form
\begin{eqnarray}
&&D_1=\frat{2~\pi^2}{\Delta^2}\left[\frat{8}{3}~\Delta^5+\Delta^3-
 \frat{8}{3}~D^5 +\frat{17}{3}~D^3-\frat{7}{2}~D+\frat{1}{2\Delta}~
\ln(\Delta+D)\right]~, \nonumber\\ &&D_2=\frat{2~\pi^2}{\Delta^2}
\left[8~D^5-8~\Delta^5+5~\Delta^3 -25~D^3+\frat{53}{2}~D-8~\frat{1}{D}
-\frat{3}{2\Delta}~\ln(\Delta+D)\right]~;\nonumber
\end{eqnarray}
The term $j~A$ of action Eq.(\ref{1}) contributes to $D_2$ as
\begin{eqnarray}
&&E_1=\frat{4~\pi^2}{\Delta^2}\left[\frat{2}{9}~D^5 -
\frat{2}{9}~\Delta^5-\frat{\Delta^3}{2}
-\frat{D^3}{18}-\frat{D}{12}-\frat{1}{12\Delta}~ \ln(\Delta+D)\right]~,
\nonumber\\
&&E_2=\frat{4~\pi^2}{\Delta^2}\left[\frat{4}{3}~\Delta^5+\frat{\Delta^3}{2}
-\frat43~D^5+\frat{17}{6}~D^3-\frat{7}{4}~D
+\frat{1}{4\Delta}~\ln(\Delta+D)\right]~;\nonumber
\end{eqnarray}
\begin{eqnarray}
&&F_1=\frat{16~\pi^2}{\Delta^2}\left[ \frat{\Delta^5}{9}-\frat{D^5}{9}+
\frat{5}{18}~D^3-\frat{D}{12} -\frat{1}{12\Delta}~ \ln(\Delta+D)\right]~,
\nonumber\\
&&F_2=\frat{16~\pi^2}{\Delta^2}\left[\frat{2}{3}~D^5- \frat{2}{3}~\Delta^5
-\frat{5}{3}~D^3+\frat{5}{4}~D-\frat{1}{2D}
 +\frat{\Delta}{4}~\ln(\Delta+D)\right]~.\nonumber
\end{eqnarray}
\begin{eqnarray}
&&H_1=\frat{4\pi^2}{\Delta^4}\left[-\frat{\Delta^7}{9}-\frat{\Delta^5}{6}
+\frat{D^7}{9}-\frat{2}{9}D^5+\frat{5}{72}D^3-\frat{D}{48}
+\left(\frat{\Delta}{12}+\frat{1}{16\Delta}\right)\ln (\Delta+D) \right]~,
\nonumber\\
&& H_2=\frat{4\pi^2}{\Delta^4}\left[-\frat{\Delta^7}{3}-\frat{\Delta^5}{6}
+\frat{D^7}{3}-D^5+\frat{75}{72}D^3-\frat{D}{16} -\left(
\frat{\Delta}{4}+\frat{5}{16\Delta}\right)\ln (\Delta+D) \right]~,\nonumber\\
&&H_3=\frat{4\pi^2}{\Delta^4}\left[\Delta^7+\frat{\Delta^5}{3}
-D^7+\frat{19}{6}D^5-\frat{85}{24}D^3+\frat{27}{16}D -\frat{5}{16\Delta}\ln
(\Delta+D) \right]~,\nonumber\\
&& H_4=\frat{4\pi^2}{\Delta^4}\left[\Delta^7-\frat{\Delta^5}{3}
-D^7+\frat{23}{6}D^5-\frat{125}{4}D^3+\frat{35}{16}D -\frat{2}{D}
+\frat{35}{16\Delta}\ln (\Delta+D) \right]~.\nonumber
\end{eqnarray}
\begin{eqnarray}
&&Y^{44}_1=8~\pi^2\left[-\frat{\Delta^5}{15}-\frat{\Delta^3}{6}-
\frat{\Delta}{8}+\frat{D^5}{15}\right]~, \nonumber\\
&&Y^{44}_2=\frat{8~\pi^2}{\Delta^2}\left[\Delta^7+\Delta^5 +\frat{\Delta^3}{8}
-D^7+\frat{5}{2}~D^5-2D^3 +\frat{D}{2}\right]~.\nonumber
\end{eqnarray}
\begin{eqnarray}
Y_1&=&\frat{16\pi^2}{\Delta^4}\left[\frat{1}{75}
\left(D^9-\Delta^9\right)-\frat{\Delta^7}{24}-\frat{\Delta^5}{24}
-\frat{11}{600}D^7 +\frat{D^5}{1200}+\frat{D^3}{960}+
\frat{D}{640}+\frat{21}{13440\Delta}\ln (\Delta+D) \right]~,\nonumber\\
 Y_2&=&\frat{16\pi^2}{\Delta^4}\left[\frat{1}{9}
\left(\Delta^9-D^9\right)+\frat{\Delta^7}{8}+\frat{\Delta^5}{48}
+\frat{13}{40}D^7 -\frat{89}{240}D^5+\frat{31}{192}D^3-
\frat{D}{128}-\frat{21}{26880\Delta}\ln (\Delta+D) \right]~,\nonumber\\
 Y_3&=&\frat{16\pi^2}{\Delta^4}\left[\frat{6}{5}
\left(D^9-\Delta^9\right)-\frat{3}{8}\Delta^7+\frat{\Delta^5}{24}
+\frat{201}{40}D^7 +\frat{1943}{240}D^5+\frat{1177}{192}D^3+\right.\nonumber\\
&+&\left. \frat{263}{128}D-\frat{1}{4D}
+\frat{21}{384\Delta}\ln (\Delta+D) \right]~.\nonumber
\end{eqnarray}
 The components of the tensors quadratic in $\delta\rho$ are the following
\begin{eqnarray}
\Sigma_1&=&\frat{4\pi^2}{\Delta^3}\left[\frat{52}{75}
\left(D^9-\Delta^9\right)-\frat23\Delta^7-\frat{184}{75}D^7
+\frat{469}{150}D^5-\frat{32}{15}D^3-\right.\nonumber\\
&-&\left.\frat{409}{120}D-\frat{17}{3D}+\frat{2}{3D^3}
+\left(\frat{367}{40\Delta}+\frat{9}{4}\Delta\right)\ln
(\Delta+D) \right]~,\nonumber\\
\Sigma_2&=&\frat{4\pi^2}{\Delta^3}\left[\frat{52}{75}
\left(D^9-\Delta^9\right)-2\Delta^7-\frat{28}{25}D^7
-\frat{77}{50}D^5+\frat{47}{10}D^3-\right.\nonumber\\
&-&\left.\frat{709}{120}D-\frat{7}{D}+\frat{1}{D^3}
+\left(\frat{367}{40\Delta}+\frat{\Delta}{12}\right)\ln
(\Delta+D) \right]~,\nonumber\\
\Sigma_3&=&\frat{4\pi^2}{\Delta^3}\left[\frat{52}{75}
\left(D^9-\Delta^9\right)-\frat23\Delta^7-\frat{184}{75}D^7
+\frat{469}{150}D^5-\frat{49}{30}D^3-\right.\nonumber\\
&-&\left.\frat{829}{120}D-\frat{2}{D}
+\left(\frat{367}{40\Delta}+\frat{49}{12}\Delta\right)\ln (\Delta+D) \right]~,
\nonumber\\
 \Sigma_4&=&\frat{4\pi^2}{\Delta^3}\left[\frat{208}{15}
\left(\Delta^9-D^9\right)+6\Delta^7+\frat{282}{5}D^7 -
\frat{441}{5}D^5+\frat{259}{4}D^3-\right.\nonumber\\
&-&\left.\frat{119}{8}D+\frat{161}{3D}-\frat{15}
 {D^3}+\frat{3}{D^5}-\frat{367}{8\Delta}\ln (\Delta+D) \right]~.\nonumber
\end{eqnarray}
The components of the tensors quadratic in $\delta\omega$ look like
\begin{eqnarray}
&&S_1=\frat{4\pi^2}{\Delta^3}\left[\frat{\Delta^7}{9}+\frat{\Delta^5}{6}
-\frat{D^7}{9}+\frat{2}{9}D^5-\frat{5}{72}D^3+\frat{D}{48}
-\left(\frat{\Delta}{12}+\frat{1}{16\Delta}\right)\ln (\Delta+D) \right]~,
\nonumber\\
&& S_2=\frat{4\pi^2}{\Delta^3}\left[-\Delta^7-\frat{\Delta^5}{3}
+D^7-\frat{19}{6}D^5+\frat{85}{24}D^3-\frat{27}{16}D +\frat{5}{16\Delta}\ln
(\Delta+D) \right]~,\nonumber\\
&&T_1=\frat{8\pi^2}{\Delta^3}\left[\frat{\Delta^9}{25}-\frat{D^9}{25}
+\frat{9}{50}D^7-\frat{63}{200}D^5+\frat{21}{80}D^3-\frat{D}{80}
-\left(\frat{\Delta}{16}+\frat{3}{40\Delta}\right)\ln (\Delta+D) \right]~,
\nonumber\\
&&T_2=\frat{8\pi^2}{\Delta^3}\left[-\frat{4\Delta^9}{5}+\frat{4D^9}{5}
-\frat{18}{5}D^7+\frat{63}{10}D^5-\frat{21}{4}D^3+\frat{15}{8}D
-\frat{1}{2D}+\frat{3}{8\Delta}\ln (\Delta+D) \right]~.\nonumber
\end{eqnarray}
\begin{eqnarray}
Z_1&=&\frat{8\pi^2}{\Delta^5}\left[\frat{29}{90}
\left(D^9-\Delta^9\right)+\frat{53}{504}\Delta^7
-\frat{3919}{2520}D^7 +\frat{523}{180}D^5-\frat{52723}{20160}D^3+\right.
\nonumber\\
&+&\left.\frat{43717}{40320}D-\frat{52}{315D}
+\left(\frat{\Delta}{32}+\frat{3}{128\Delta}\right)\ln (\Delta+D) \right]~,
\nonumber\\
Z_2&=&\frat{8\pi^2}{\Delta^5}\left[\frat{17}{45} \left(D^9-\Delta^9\right)
+\frat{11}{504}\Delta^7 -\frat{4339}{2520}D^7 +\frat{2197}{720}D^5-
\frat{53633}{20160}D^3+\right.\nonumber\\
&+&\left.\frat{44137}{40320}D-\frat{52}{315D}
+\left(\frat{\Delta^3}{24}+\frat{5}{32}\Delta+
\frat{3}{128\Delta}\right)\ln (\Delta+D) \right]~,\nonumber\\
Z_3&=&\frat{8\pi^2}{\Delta^5}\left[\frat{233}{90}
\left(\Delta^9-D^9\right)-\frat{29}{72}\Delta^7 +\frat{4339}{360}D^7
-\frat{7847}{360}D^5+\frat{54703}{2880}D^3-\right.\nonumber\\
 &-&\left.\frat{44077}{5760}D+\frat{52}{45D}
-\left(\frat{5}{32}\Delta+\frat{21}{128\Delta}\right)
\ln (\Delta+D) \right]~,\nonumber\\
Z_4&=&\frat{8\pi^2}{\Delta^5}
\left[\frat{94}{45} \left(\Delta^9-D^9\right)-\frat{41}{72}\Delta^7
+\frat{3589}{360}D^7 -\frat{13279}{720}D^5+\frat{44173}{2880}D^3-\right.
\nonumber\\
&-&\left.\frat{39217}{5760}D+\frat{52}{45D}
-\frat{21}{128\Delta}\ln (\Delta+D) \right]~,\nonumber\\
Z_5&=&\frat{8\pi^2}{\Delta^5}\left[\frat{114}{5}
\left(D^9-\Delta^9\right)+\frat{33}{8}\Delta^7 -\frat{4269}{40}D^7
+\frat{1519}{80}D^5-\frat{54213}{320}D^3+\right.\nonumber\\
&+&\left.\frat{44337}{640}D-\frat{57}{5D} +\frat{189}{128\Delta}\ln
(\Delta+D) \right]~.\nonumber
\end{eqnarray}
\begin{eqnarray}
\Phi^{44}_1&=&\frat{8\pi^2}{\Delta^3}\left[\frat{1}{25}
\left(\Delta^9-D^9\right)+\frat{\Delta^7}{12}+\frat{\Delta^5}{24}
-\frat{29}{300}D^7 -\frat{13}{200}D^5-\frat{D^3}{480}+
\frat{D}{320}+\frat{1}{320\Delta}\ln (\Delta+D) \right]~,\nonumber\\
\Phi^{44}_2&=&\frat{8\pi^2}{\Delta^3}\left[\frat{4}{5}
\left(D^9-\Delta^9\right)-\frat{3}{4}\Delta^7-\frat{\Delta^5}{12}
 -\frat{57}{20}D^7 +\frat{451}{120}D^5-\frat{209}{96}D^3+
\frat{31}{64}D-\frat{1}{64\Delta}\ln (\Delta+D) \right]~,\nonumber
\end{eqnarray}
\begin{eqnarray}
\Phi_1&=&\frat{16\pi^2}{\Delta^5}\left[\frat{11}{150}
\left(\Delta^{11}-D^{11}\right)+\frat{\Delta^9}{60}+\frat{\Delta^7}{96}
+\frat{D^9}{50}- \frat{43}{2400}D^7+\frat{7}{4800}D^5+\right.\nonumber\\
&+&\left. \frat{3}{1280}D^3-\frat{3}{2560}D
+\left(\frat{\Delta}{320}+\frat{1}{512\Delta}\right)\ln
(\Delta+D) \right]~,\nonumber\\
\Phi_2&=&\frat{16\pi^2}{\Delta^5}\left[\frat{11}{15}
\left(D^{11}-\Delta^{11}\right)-\frat{3}{40}\Delta^9-\frat{\Delta^7}{96}
-\frat{7}{24}D^9+ \frat{239}{48}D^7-\frat{49}{120}D^5+\right.\nonumber\\
&+&\left.\frat{119}{768}D^3-\frat{3}{512}D
-\left(\frat{\Delta}{128}+\frat{7}{512\Delta}\right)\ln (\Delta+D) \right]~,
\nonumber\\
\Phi_3&=&\frat{16\pi^2}{\Delta^5}\left[
\Delta^{11}-D^{11}+\frat{3}{10}\Delta^9-\frat{\Delta^7}{32}
+\frat{52}{10}D^9- \frat{1759}{160}D^7+\frat{3829}{320}D^5-\right.\nonumber\\
&-&\left.\frat{1771}{256}D^3-\frat{959}{512}D-\frat{1}{4D}
+\frat{63}{512\Delta}\ln (\Delta+D) \right]~.\nonumber
\end{eqnarray}
The components of kinetic energy tensors which are the cross terms between
$\delta\rho$ and $\delta\omega$ are defined as follows
\begin{eqnarray}
&&\Psi_1=\frat{4\pi^2}{\Delta^3}\left[\frat{16}{9}\Delta^7
-\frat{16}{9}D^7+\frat{56}{9}D^5-\frat{70}{9}D^3+\frat{23}{6}D
-\frat{2}{D}-\left(\frat{\Delta}{3}-\frat{3}{2\Delta}\right)\ln
 (\Delta+D) \right]~,\nonumber\\
&& \Psi_2=\frat{4\pi^2}{\Delta^3}\left[\frat{8}{9}\Delta^7-\frat{8}{9}D^7
+\frat{28}{9}D^5-\frat{35}{9}D^3+\frat{D}{6}+
\left(\frat{4\Delta}{3}+\frat{3}{2\Delta}\right)\ln (\Delta+D) \right]~,
\nonumber\\
 &&\Psi_3=\frat{4\pi^2}{\Delta^3}\left[-8\Delta^7+8D^7
-28D^5+35D^3-\frat{31}{2}D+\frat{8}{D} -\frat{15}{2\Delta}\ln (\Delta+D)
\right]~,\nonumber
\end{eqnarray}
\begin{eqnarray}
&&W_1=\frat{16\pi^2}{\Delta^3}\left[\frat{2}{25}(D^9-\Delta^9)
-\frat{9}{25}D^7+\frat{63}{100}D^5-\frat{21}{40}D^3+\frat{2}{5}D
-\left(\frat{\Delta}{8}+\frat{9}{40\Delta}\right)\ln (\Delta+D) \right]~,
\nonumber\\
&&W_2=\frat{16\pi^2}{\Delta^3}\left[\frat{8}{5}(\Delta^9-D^9)
+\frat{36}{5}D^7-\frat{63}{5}D^5+\frat{21}{2}D^3-\frat{33}{8}D
-\frat{3}{4D}+\frat{1}{4D^3}+\frat{9}{8\Delta}\ln (\Delta+D) \right]~,
\nonumber
\end{eqnarray}
\begin{eqnarray}
&&X_1=\frat{8\pi^2}{\Delta^3}\left[\frat{D^7}{18}-\frat{\Delta^7}{18}
 -\frat{7}{36}D^5+\frat{35}{144}D^3-\frat{D}{96}
-\left(\frat{\Delta}{12}+\frat{3}{32\Delta}\right)\ln (\Delta+D) \right]~,
\nonumber\\
&&X_2=\frat{8\pi^2}{\Delta^3}\left[\frat{\Delta^7}{2}
-\frat{D^7}{2}+\frat{7}{4}D^5-\frat{35}{16}D^3+\frat{31}{32}D
-\frat{1}{2D}+\frat{15}{32\Delta}\ln (\Delta+D) \right].\nonumber
\end{eqnarray}
\begin{eqnarray}
 \Theta^{44}_1&=&\frat{48\pi^2}{\Delta^3}\left[\frat{1}{150}
\left(\Delta^{9}-D^{9}\right)+\frat{3}{100}D^7-
\frat{21}{400}D^5+\frat{7}{160}D^3-\frat{D}{480}-
\left(\frat{\Delta}{96}+ \frat{1}{80\Delta}\right)\ln (\Delta+D) \right]~,
\nonumber\\
\Theta^{44}_2&=&\frat{48\pi^2}{\Delta^3}\left[\frat{2}{15}
\left(D^9 -\Delta^9\right)-\frat{3}{5}D^7+
\frat{21}{20}D^5-\frat{7}{8}D^3+\frat{5}{16}D-
\frat{1}{12D}+ \frat{1}{16\Delta}\ln (\Delta+D) \right]~,\nonumber
\end{eqnarray}
\begin{eqnarray}
\Theta_1&=&\frat{96\pi^2}{\Delta^5}\left[\frat{1}{900}
\left(\Delta^{11}-D^{11}\right)+\frat{11}{1800}D^9-
\frat{11}{800}D^7+\frat{77}{4800}D^5-\frat{D^3}{640}+\right.\nonumber\\
&+&\left.\frat{D}{1280} -\left(\frat{\Delta^3}{192}+\frat{\Delta}{80}+
\frat{5}{768\Delta}\right)\ln (\Delta+D) \right]~,\nonumber\\
\Theta_2&=&\frat{96\pi^2}{\Delta^5}\left[\frat{1}{90}
\left(D^{11}-\Delta^{11}\right)-\frat{11}{180}D^9+
\frat{11}{80}D^7-\frat{77}{480}D^5+\frat{37}{384}D^3-\right.\nonumber\\
&-&\left.\frat{53}{768}D +\left(\frat{\Delta}{32}+ \frat{35}{768\Delta}
\right)\ln
(\Delta+D) \right]~,\nonumber\\ \Theta_3&=&\frat{96\pi^2}{\Delta^5}
\left[\frat{1}{6}
\left(\Delta^{11}-D^{11}\right)+\frat{11}{12}D^9-
\frat{33}{16}D^7+\frat{77}{32}D^5-\frat{195}{128}D^3+\right.\nonumber\\
 &+&\left.\frat{151}{256}D+\frat{7}{24D}-\frat{1}{24 D^3}
-\frat{315}{768\Delta}\ln (\Delta+D) \right]~.\nonumber
 \end{eqnarray}
 \begin{eqnarray}
 \Xi_1&=&\frat{32\pi^2}{\Delta^5}\left[\frat{29}{180}
\left(\Delta^9-D^9\right)+\frat{29}{40}D^7
-\frat{203}{160}D^5+\frat{20497}{20160}D^3-\frat{12}{35}D-\right.\nonumber\\
&-&\left.\frat{2}{35D}+\frat{13}{315D^3}
+\frat{3}{64}\left(\Delta+\frat{1}{\Delta}\right)\ln (\Delta+D) \right]~,
\nonumber\\
\Xi_2&=&\frat{32\pi^2}{\Delta^5}\left[\frat{17}{90}
\left(\Delta^9-D^9\right)+\frat{17}{20}D^7
-\frat{119}{80}D^5+\frat{1921}{1680}D^3-\frat{2339}{6720}D-\right.\nonumber\\
&-&\left.\frat{2}{35D}+\frat{13}{315D^3}
+\left(\frat{\Delta^3}{24}+\frat{3}{32}\Delta+ \frat{3}{64\Delta}\right)
\ln (\Delta+D) \right]~,\nonumber\\ \Xi_3&=&\frat{32\pi^2}
{\Delta^5}\left[\frat{233}{180} \left(D^9-\Delta^9\right)-\frat{233}{40}D^7
 +\frat{1631}{160}D^5-\frat{23767}{2880}D^3+\frat{449}{160}D+\right.
\nonumber\\
&+&\left.\frat{2}{5D}-\frat{13}{45D^3} -\left(\frat{15}{64}\Delta+
\frat{21}{64\Delta}\right)\ln (\Delta+D) \right]~,\nonumber\\
 \Xi_4&=&\frat{32\pi^2}{\Delta^5}\left[\frat{47}{45} \left(D^9-\Delta^9\right)
-\frat{47}{10}D^7 +\frat{329}{40}D^5-\frat{9611}{1440}D^3
+\frat{663}{320}D+\right.
\nonumber\\
&+&\left.\frat{13}{20D}-\frat{13}{45D^3}
-\frat{21}{64\Delta}\ln (\Delta+D) \right]~,\nonumber\\
\Xi_5&=&\frat{32\pi^2}{\Delta^5}\left[\frat{57}{5}
\left(\Delta^9-D^9\right)+\frat{513}{10}D^7
-\frat{3591}{40}D^5+\frat{11691}{160}D^3-\frat{7407}{320}D-\right.\nonumber\\
&-&\left.\frat{117}{20D}+\frat{57}{20D^3}
+\frat{189}{64\Delta}\ln (\Delta+D) \right]~.\nonumber
\end{eqnarray}
The tensors which are not integrated in the elementary functions read
\begin{eqnarray}
&&B_i=8~\int dy~ \frat{y_i}{|{\vf y} + {\vf \Delta}|^2}~\frat{2~y^2-
{\vf y}^2}{y^2~Y^3}~,~~C_{ijk}=8~\int dy~
\frat{y_i~y_j~y_k}{|{\vf y} + {\vf \Delta}|^2}~\frat{1}{y^2~Y^3}~,\nonumber\\
&&O_i=4~\int dy~ \frat{y_i}{|{\vf y} +
{\vf \Delta}|^2}~\frat{1}{y^2~Y^2}~,~~~~~~
 P_{ijk}=4~\int dy~ \frat{y_i~y_j~y_k}{|{\vf y} +
 {\vf \Delta}|^2}~\frat{1}{y^2~Y^2}~,\nonumber\\
&& Q_{ij}=12 \int dy\frat{y_i y_j}{|{\vf y} +
{\vf \Delta}|^2} \frat{y^2-1}{Y^4}~,~~~~~~
R_{ijkl}=4\int dy\frat{y_i y_j y_k y_l}
{|{\vf y}+{\vf\Delta}|^2}\left(\frat{1}{y^2Y^3}+\frat{2}{Y^4}-\frat{4}
{y^2Y^4}\right)~, \nonumber\\
&&U_{ij}=2\int dy\frat{y_i y_j}{|{\vf y} + {\vf \Delta}|^2}\frat{1}{y^2Y^2}~,
~~~~~~~~~V_{ijkl}=2\int dy\frat{y_i y_j y_k y_l}{|{\vf y} + {\vf \Delta}|^2}
\frat{1}{y^2~Y^2}~,\nonumber
\end{eqnarray}
where $Y=y^2+1$. The coefficients of the obtained polynomials in
$\Delta$ and $D$ should be rather peculiarly organized.
Because at large distances the corresponding integrals  are the
decreasing functions.
Therefore the coefficients at high powers should
be tuned in such a way to cancel each other, up to the order necessary
for functions decreasing, at
expanding
$D=\sqrt{\Delta^2+1}$, when $\Delta\to\infty$. At small $\Delta$ the
singularity
of terms related to $\sim \frat{\ln \Delta}{\Delta}$ what means the sum of all
the coefficients, excepting the terms proportional to $\sim \Delta^3~ \ln
\Delta$,
 $\Delta~\ln \Delta$ and the powers of $\Delta$, should equal to zero.


\end{document}